\documentclass{appolb}
\usepackage{epsfig}
\def\bea{\begin{eqnarray}}
\def\eea{\end{eqnarray}}
\def\be{\begin{equation}}
\def\ee{\end{equation}}
\newcommand{\anp}{Ann.\ Phys.\ }

\begin{document}
\title{Manipulating the shape of electronic non-dispersive wave-packets in
the hydrogen atom: numerical tests in realistic experimental conditions}
\author{Dominique Delande
\address{Laboratoire Kastler Brossel, Universit\'e Pierre et Marie Curie\\
Case 74, 4 place Jussieu, 75252 Paris, France}
\and
Krzysztof Sacha and Jakub Zakrzewski
\address{
 Instytut Fizyki imienia Mariana Smoluchowskiego,
  Uniwersytet Jagiello\'nski,\\
 ulica Reymonta 4, PL-30-059 Krak\'ow, Poland}
}

\date{June, 4, 2002}
\maketitle
\begin{abstract}
We show that combination of a linearly polarized resonant 
microwave field and a parallel static
electric field may be used to create a non-dispersive electronic wavepacket
in Rydberg atoms. The static electric field allows for manipulation of the shape
of the elliptical trajectory the wavepacket is propagating on. 
Exact quantum numerical calculations for realistic experimental
parameters show that the wavepacket evolving 
on a linear orbit can be very easily prepared in a laboratory either by a direct 
optical excitation or by preparing an atom in an extremal Stark state 
and then slowly 
switching on the microwave field. The latter scheme seems to be very resistant 
to experimental imperfections.  
Once the wavepacket on the linear orbit is excited, the static
field may be used to manipulate the shape of the orbit.
\end{abstract}

\PACS{PACS: 05.45.+b, 32.80.Rm, 42.50.Hz}

\section{Introduction}
Atomic wavepackets form a bridge which allows to understand the mutual
relations between the classical and the quantum world \cite{ber+98,alb+91,BDZ02}.
Typically (i.e. with the notable exception of the harmonic oscillator)
any initially localized wavepacket (which mimics a classical particle)
will have its center of mass follow a classical trajectory at short time, 
but will progressively
spread in time. Recent years have brought several attempts to overcome the
spreading.

The wavepacket spreading is classical in character and is due to the nonlinearity
of the Hamiltonian, or saying it differently, the fact that trajectories
with different energies evolve with different frequencies \cite{BDZ02}.
 To overcome the 
classical spreading of a bunch of particles, the phenomenon of 
{\em nonlinear resonance} can be used (e.g while guiding
particles in accelerators). The idea is very simple:
a classical nonlinear system periodically driven by an external
perturbation displays resonances when the period of the external 
driving matches the period of the unperturbed motion. In a region of phase
space called  nonlinear resonance island, the internal
motion of the system becomes locked on the external drive. 
This motion in the resonance island is similar to pendulum oscillations.
The corresponding quantum description was first given by Berman and Zaslavsky~\cite{zas} 
and later readressed by
Henkel and Holthaus~\cite{hen+92,hol95}, who also gave
the semiclassical interpretation.
For realistic systems, first
studies involved the hydrogen atom driven by either a 
linearly~\cite{del+94,buch+95} or 
circularly~\cite{ibb,dzb95,zdb95} polarized microwave field.
 In the latter situation, a
particularly simple picture may be obtained since, in the frame rotating with
the microwave electric field, the Hamiltonian becomes time
independent and the center of the pendulum island becomes a fixed point of
the dynamics. This allows for an approximate description 
of the states 
localized near the stable fixed point using Gaussian
wavepackets, termed then Trojan states~\cite{ibb}.
 
Several properties of nondispersive wavepackets have been analysed recently.
The mixed semiclassical/quantum description is particularly convenient.
Because of the time periodicity of the driving, the Floquet theorem~\cite{s65}
can be used: it ensures the existence of a basis of states (the eigenstates of the
Floquet Hamiltonian) which evolve periodically in time, and that any solution
of the time-dependent Schr\"odinger equation can be expanded in a simple
linear combination of these quasienergy eigenstates. If a 
single quasienergy
eigenstate is initially localized, it will preserve this localization
during the temporal evolution or, more precisely, recover it
every period of the drive~\cite{dzb95}, and thus overcome the
long-time spreading. From semiclassics, such states can be constructed
as  localized inside
the nonlinear island and thus constitute nondispersive wavepackets
locked on the external driving. Because they are built from two robust structures
(the classical resonance island and the basis given by the
Floquet theorem), the nondispersive wavepackets are robust versus any small
perturbation not taken into account in the preceding approach.
For example, for microwave driven atoms, the
nondispersive wavepacket states are resistant to any geometrical
imperfection in the field direction, amplitude or polarization,
and protected from direct fast ionization by the resonance island
(they can however ionize after tunneling outside the island and their 
lifetimes exhibit interesting
fluctuations \cite{zdb95,zdb98}). Their decay due to spontaneous emission
of photons  has
also been analysed \cite{zibb97,dz98,horn98}.

These wavepacket states  have most probably already been prepared in experiments studying
the microwave induced ionization of hydrogen atom. Indeed, atoms were found 
to be relatively stable against ionization for
microwave frequency close to Kepler frequency~\cite{bay+89,gal+88,bel+96,noe00}.
However,
those experiments were addressing different issues and, most probably,
populated several Floquet states at once. For an unambiguous identification
of  nondispersive wavepackets, special preparation (as well as
detection) schemes have to be envisioned. In fact, for wavepackets
driven by a circularly polarized microwave field,
such a scheme has been proposed quite early \cite{dzb95}.
It requires the preparation of an atom in an initial Rydberg circular state, 
followed by a slow turn on of the
circularly polarized microwave. This scheme has been simulated numerically
for realistic parameters~\cite{zd97} but, up till now, no experimental test
has been made.
 
It would be desirable to simplify the proposed scheme, especially
to avoid the initial preparation of a circular state, which is possible
but not trivial. By contrast, it is much simpler to excite
a Rydberg state where the electron probability near the nucleus
is important: a direct optical excitation from the ground state
or a low excited state is simple and convenient. Moreover, because of
the monochromatic character of the laser sources, it is rather simple to
excite {\em selectively} the desired Rydberg state and not its neighbors.
Thus, the simplest idea
is to change the classical electron trajectory (on which the nondispersive
wavepacket is built) to
an elongated Kepler orbit hitting the nucleus. Such a nondispersive
wavepacket can be easily achieved using a resonant
microwave field linearly polarized along the degenerate Kepler orbit.
There is however a undesirable side effect: the motion along the polarization axis is
transversally unstable, that is any deviation from strict alignment of the
electron along the microwave field will be exponentially amplified with time.
This problem may be overcome by addition of a static electric
field, parallel to the polarization axis as shown by us using a semiclassical
approach~\cite{szd98}. The resulting situation 
is very attractive from the experimental point 
of view. In~\cite{szd98}, we proposed two possible experimental schemes:
either the direct optical excitation of the ``linear" nondispersive
wavepackets in the presence of both the linearly polarized microwave field
and the stabilizing static field (scheme I) or the
excitation of a convenient Stark state (in the presence of the
static field only) followed by an adiabatic turn on of the
microwave field (scheme II). Scheme I is simpler, but might
be less convenient in a real experiment because it
requires that the laser beam is sent inside the microwave cavity,
which may be difficult. Scheme II requires some control
on how the microwave field is turned on.
Once the ``linear" nondispersive wavepacket is excited by either of the
two schemes, a subsequent decrease of the {\em static} field can lead
to an ``elliptical" nondispersive wavepacket, i.e. localized on
an elliptical Kepler trajectory of arbitrary eccentricity and, in the
limit of vanishing static field to a ``circular" nondispersive wavepacket.
The aim of this paper is to explore 
the feasibility of such schemes in a real experiment. We will use realistic
parameters and the combination of a semiclassical approach (in order to get the
orders of magnitudes and a physical picture of what is going on) and of full
quantum numerical simulations (which provide accurate numbers).
Such an analysis reveals also possible experimental difficulties
overlooked in the semiclassical discussion.

A very
similar experiment has been performed by Bromage and Stroud \cite{bs99} who, 
starting from an
extremal Stark state of the sodium atom, excited a wavepacket by applying a short
electromagnetic half-cycle pulse which localized an electron on a highly
eccentric orbit. After excitation, there was no mechanism to prevent the  
wavepacket from spreading. Nevertheless, 
the authors observed a few nice oscillations in the ionization 
signal reflecting the classical motion of the electron.
Thus, it seems that only one step further is needed to obtain a nondispersive
wavepacket in the laboratory. Namely the short pulse excitation 
should be replaced by a slow turn on of the microwave field whose presence  
afterwards assures the nonspreading character of the created wavepacket.

\section{The quasi-energy spectrum at fixed static and microwave fields:
confrontation of semiclassical and quantum results}

We consider an hydrogen atom exposed to both a static electric field and
a linearly polarized microwave field parallel to the static field. 
The Hamiltonian of
the system reads (in atomic units, with the fields along the $z$ axis): 
\be
H=\frac{{\mathbf p}^2}{2}-\frac{1}{r}+Fz\cos(\omega t)+F_sz,
\label{ham}
\ee
where $F$ and $\omega$ stand for the amplitude and frequency of the microwave field
respectively,
while $F_s$ is the amplitude of the static field. The system is invariant under
rotation around the $z$ axis, and the angular momentum projection on
this axis is consequently conserved. In the following,
we will assume for simplicity $L_z=0.$
Similar conclusions can be reached for low values
of $L_z.$ 

The Hamiltonian (\ref{ham}) is
time-periodic; the Floquet theorem \cite{s65} 
implies that any 
solution of the
Schr\"odinger equation can be written as a linear combination of the Floquet
eigenstates. Those are time-periodic (with period $2\pi/\omega$) 
eigenfunctions $|\phi_\alpha(t)\rangle$ of the so-called Floquet Hamiltonian operator
\be
{\cal H}|\phi_\alpha(t)\rangle =\left(H-i\frac{\partial}{\partial t}\right)
|\phi_\alpha(t)\rangle=\varepsilon_\alpha \ |\phi_\alpha(t)\rangle,
\label{floquet}
\ee
where $\varepsilon_\alpha$ are the quasienergies of the system. Thus the
preparation of an atom in a single Floquet state ensures
that the electronic density evolves periodically in time. However
not every eigenstate $|\phi_\alpha(t)\rangle$ corresponds to a well localized electron
propagating on a classical trajectory.
To find which Floquet states are the nondispersive wavepackets, we
need a semiclassical approach. In order to study the classical dynamics
of such a time-dependent system, we need to define the extended
phase space \cite{lil}
where one deals with the additional momentum $P_t$ conjugate to the $t$ (time)
variable.
The temporal evolution is described by the 
Hamiltonian function ${\cal H}=H+P_t,$ which
is the  classical analog of the quantum Floquet
operator defined in eq.~(\ref{floquet}). 

Consider a hydrogen atom illuminated by a microwave field
of frequency:
\be 
\omega = \frac{1}{n_0^3}.
\ee
$n_0$ is the effective principal quantum number which is resonant with
the external driving, that is such that the unperturbed
Kepler motion has the frequency $\omega_K=\omega$ (in the classical
language) or such that the microwave perturbation is almost
resonant with
the transitions to the $n_0'=n_0\pm 1$ neighboring states (in the
quantum language).

At large microwave field, the classical phase space structure may be
extremely complicated with interleaved regions of chaotic and regular
motion. At relatively small microwave field -- the situation we are interested in --
the resonance between 
the driving frequency and the frequency of the unperturbed motion leads to 
a strong perturbation of the system and the creation of a stable island in 
phase space centered on a periodic orbit at exactly the frequency $\omega$.
There, the effect of non resonant term (which are responsible for the
onset of chaos at strong field) can be safely neglected. In this so-called
secular approximation, it is assumed that the motion in the resonance
island is much slower than the Kepler motion itself. The Hamiltonian 
can be rewritten in the unperturbed action-angle coordinates 
$(I,\theta,L,\psi)$ which 
describe the classical Kepler motion. The total action $I$ is the classical
equivalent of the principal quantum number (so that the Hamiltonian of the
unperturbed hydrogen atom is $-1/2I^2.)$ The conjugate angle $\theta$ describes
the motion along the classical Kepler orbit (it evolves periodically at a constant
angular velocity $\omega_K$). The other pair of action-angle
coordinates $(L,\psi)$ describes the parameters of the classical Kepler
ellipse, i.e. the total angular momentum $L$ related to the eccentricity
by
\be
e = \sqrt{1-\frac{L^2}{I^2}},
\ee 
and the conjugate angle $\psi$ between the major axis of the classical Kepler
ellipse and the field axis.
A convenient approach is to
switch to the {\it rotating} frame defined by:
\begin{eqnarray}
\hat{\theta} = \theta-\omega t\\
\hat{P}_t = P_t + \omega I
\end{eqnarray}
because $\hat{\theta}$ appears as a slowly varying variable. The secular approximation
is to neglect all terms in the Hamiltonian which oscillate at the microwave frequency
or its harmonics.
The effective
Hamiltonian function describing the motion in the stable resonant island 
(for details see \cite{BDZ02,szd98,szd99b}) thus reads:

\begin{eqnarray}
{\cal H}_{\mathrm{sec}}&=&\hat{P}_t-\frac{1}{2I^2}-\omega I 
 - \frac{3eF_sI^2}{2}\cos\psi \nonumber\\
&+&FI^2\left\{-J'_1(e)\cos\psi\cos\hat{\theta} + 
\frac{\sqrt{1-e^2}J_1(e)}{e} \sin\psi \sin\hat{\theta}\right\} 
\label{hsec}
\end{eqnarray}
where $J_1$ and $J'_1$ denotes the Bessel function and its derivative, respectively. 

Having the effective Hamiltonian, the last stage is to quantize the system.
The radial motion, i.e. in the $(I,\hat{\theta})$ space, effectively decouples from 
the slow angular 
motion in the $(L,\psi)$ space. In effect, one can quantize the
system in the spirit of the Born-Oppenheimer approximation, i.e. first quantize 
the radial motion keeping the secular motion frozen \cite{bsdz98,szd99b}, using 
the results to construct an effective potential for the motion in the
$(L,\psi)$ space. In the limiting case when the motion
in the $(I,\hat{\theta})$ space is harmonic such a procedure was followed in \cite{szd98}.
 This is however
not suitable for very low microwave fields. We give in the appendix the 
derivation and results in the general case. 
It should also be noted that, because the Coulomb potential is an homogeneous 
function (of degree $-1$) of the position while both the static and
the microwave field interaction Hamiltonians are homogeneous functions
of degree 1, there exist a classical scaling invariance law which allows
to express the classical dynamics with the scaled quantities:
\bea
F_0&=&Fn_0^4, \\ 
F_{s,0}&=&F_sn_0^4, \\
L_0&=&\frac{L}{n_0}.
\eea

We have chosen to perform all numerical calculations (semiclassical and quantum)
for $n_0=60.$ This value corresponds to the principal quantum number of
Rydberg states prepared in a typical experiment. The corresponding resonant
microwave frequency is $1/(60)^3$ in atomic units, i.e. 30.48 GHz. This is
in the microwave regime where efficient high-quality sources are available.
The electric field amplitude such that $F_0 = 0.01$ 
(a typical value to be used in an experiment, see below)
is $0.01/(60)^4$ in atomic units,
i.e. 4 V/cm. Producing a static or microwave field with such an amplitude
is not a problem in a real experiment.

\begin{figure}
\begin{center}
\includegraphics*[width=9cm]{sclE.eps}
\includegraphics*[width=9cm]{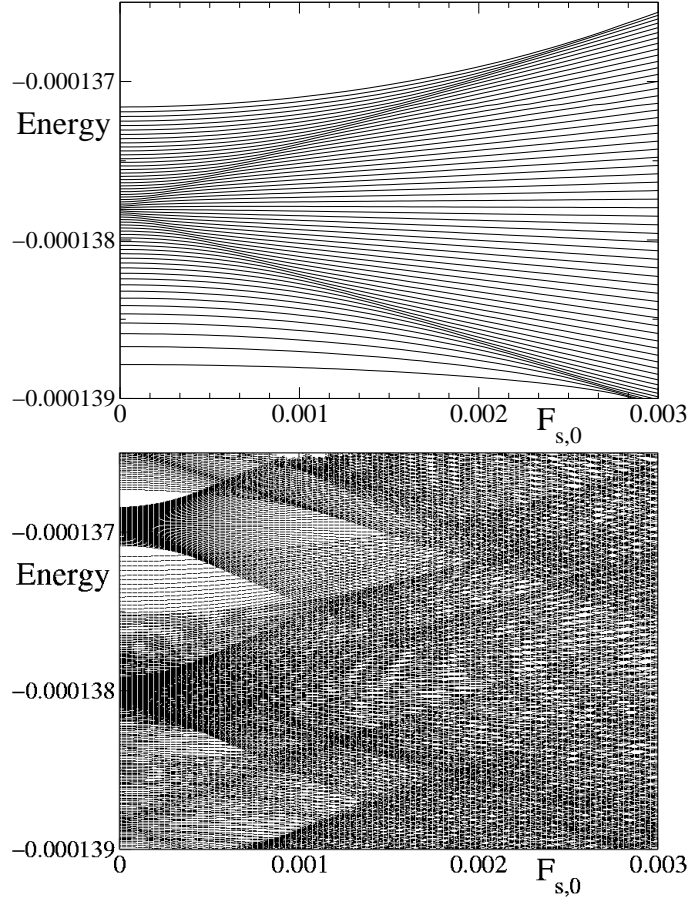}
\end{center}
\caption{Quasienergy levels of the hydrogen atom exposed to parallel
 static and microwave fields as a function of the scaled static field $F_{s,0}$ 
 for $n_0=60$ (microwave frequency 
30.48 GHz) and fixed scaled microwave field amplitude $F_0=0.015$ (i.e. 6V/cm).
The upper panel shows the levels of the $n=60$ manifold
calculated using a semiclassical approach. The highest level
of the manifold is the non-dispersive electronic wavepacket. For $F_{s,0}=0,$ it propagates along a circular
trajectory. At increasing $F_{s,0},$ it smoothly turns into an elliptical non-dispersive wavepacket
 -- that is a wavepacket propagating along an elliptical trajectory -- and finally to a linear
 wavepacket propagating along the field axis. The lower panel shows the quasienergy levels obtained
 from an exact numerical diagonalization of the Floquet Hamiltonian. The levels belonging to the
 $n=60$ manifold almost coincide with the semiclassical prediction.}
\label{levdyn2}
\end{figure}  
    
In Fig.~\ref{levdyn2}, we show the quasienergy levels of the 60 states
belonging to the resonant $n=60$ manifold as a function of the
scaled static field $F_{s,0},$ for a fixed scaled microwave amplitude
$F_0=0.015.$ For this calculation, we used the semiclassical approach
described in the appendix. It is essentially identical to Fig.~2 of \cite{szd98},
for slightly different field values, but the general Mathieu quantization 
(described in the appendix) was used instead of the harmonic approximation
used in \cite{szd98}. 
The purely quantum quasienergy spectrum -- obtained from numerical
diagonalization of the Floquet Hamiltonian -- is also shown in Fig.~\ref{levdyn2}.
It is remarkably similar to the semiclassical spectrum for the $n=60$ manifold.
However, the total quantum spectrum is very congested, with plenty
of other manifolds superimposed. These manifolds correspond to lower or
higher principal quantum numbers shifted (in energy) by an integer multiple
of the microwave frequency $\omega.$ It happens that, by chance, some of these
manifolds overlap with the $n=60$ manifold. The most striking result is that
these manifolds overlap but (almost) do not interact! Indeed, a careful inspection
shows that there are no level crossings but rather very small avoided
crossings (invisible at the scale of the figure). This is easily understood
from the (semi)classical dynamics. Indeed, the nonlinear resonance island
isolates the non-dispersive wavepackets from other states localized outside
the island. In quantum mechanics, they are coupled only by tunneling,
a typically very weak process responsible for the tiny avoided crossings.

\begin{figure}
\begin{center}
\includegraphics[width=5cm,angle=-90]{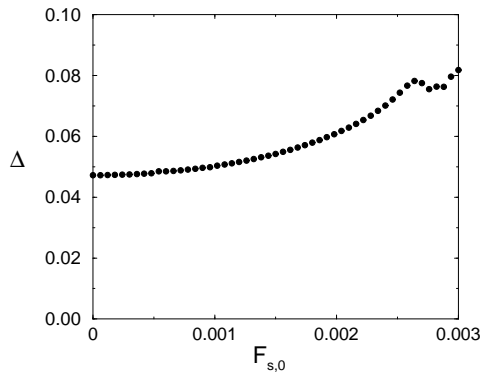}
\end{center}
\caption{Difference between the semiclassical prediction and
quantum results for the energy
of the non-dispersive wavepacket of the hydrogen atom in the presence of
parallel microwave and static electric field, as a function of the scaled
static field amplitude $F_{s,0}$. The parameters are those of figure
\protect{\ref{levdyn2}}, that is $n_0=60, F_0=0.015$. The energy difference $\Delta$
is plotted in units of the mean level spacing of the Floquet Hamiltonian,
estimated to be equal to $2/n_0^4.$ $\Delta$ is much smaller than one, which
proves the high-quality of the semiclassical approximation.}
\label{energy_difference}
\end{figure}

The usefulness of the semiclassical approach must be emphasized. Without
the guideline it provides, it would be impossible to recognize the
$n=60$ manifold and identify the non-dispersive wavepackets in the
mess of lower panel in Fig.~\ref{levdyn2}. 
A further test of the accuracy of the
semiclassical approximation is provided by a direct comparison of the
numerical prediction for the quasienergy levels with the exact quantum levels.
The difference is plotted in Fig.~\ref{energy_difference} for the upper
state of the $n=60$ manifold, as a function of the scaled static field.
The energy difference is plotted in units of the mean level spacing,
which is here of the order of $2/n_0^4$~\cite{dzb95}. For all field
values, it is smaller than 10\% of the mean level spacing and
it evolves essentially smoothly with the field. This implies that
the semiclassical approximation catches most of the physics of the system.
It also means that it can be used to easily find a state of interest
among all energy levels obtained from a numerical diagonalization. 
More importantly, for $n_0=60,$ the energy difference between the semiclassical 
and quantum results is about $10^{-8}$ atomic units,
corresponding to a frequency difference of 60 MHz. In a real
experiment, the semiclassical prediction will thus gives a very useful
indication for exciting the right spectral line.

In zero static field (left of  figure~\ref{levdyn2}), one can see
the manifold of states created in the presence of the microwave field only,
with low-energy states associated with the weakest interaction
with the microwave and consequently to the worst localization along
the Kepler orbit (the latter being orbits mainly perpendicular to the
field direction). In the middle of the manifold, one can see a local
minimum spacing associated with the hyperbolic point at $(L=0,\psi=0)$, i.e.
the degenerate linear Kepler orbit along the microwave field axis. As mentioned
above, this motion is transversally unstable. The corresponding
Floquet eigenstates are thus well localized longitudinally (they form
nice wavepackets propagating back and forth) but poorly localized angularly.
At the top of the manifold, there are states with maximum longitudinal localization
and trajectories close to circular. Note, however, that these are
$L_z=0$ states and that the circular trajectory is in the plane containing
the quantization axis, so that the total wavefunction is localized on a sphere
in the three-dimensional space, in sharp contrast with the so-called
``circular" states (for pure hydrogen) which are localized on a circle
 perpendicular to the quantization axis or their combinations building
 Trojan-like \cite{ibb,dzb95} wavepackets for circularly polarized microwave.

\begin{figure}
  \begin{center}
 \includegraphics[width=0.4\textwidth, clip = true]{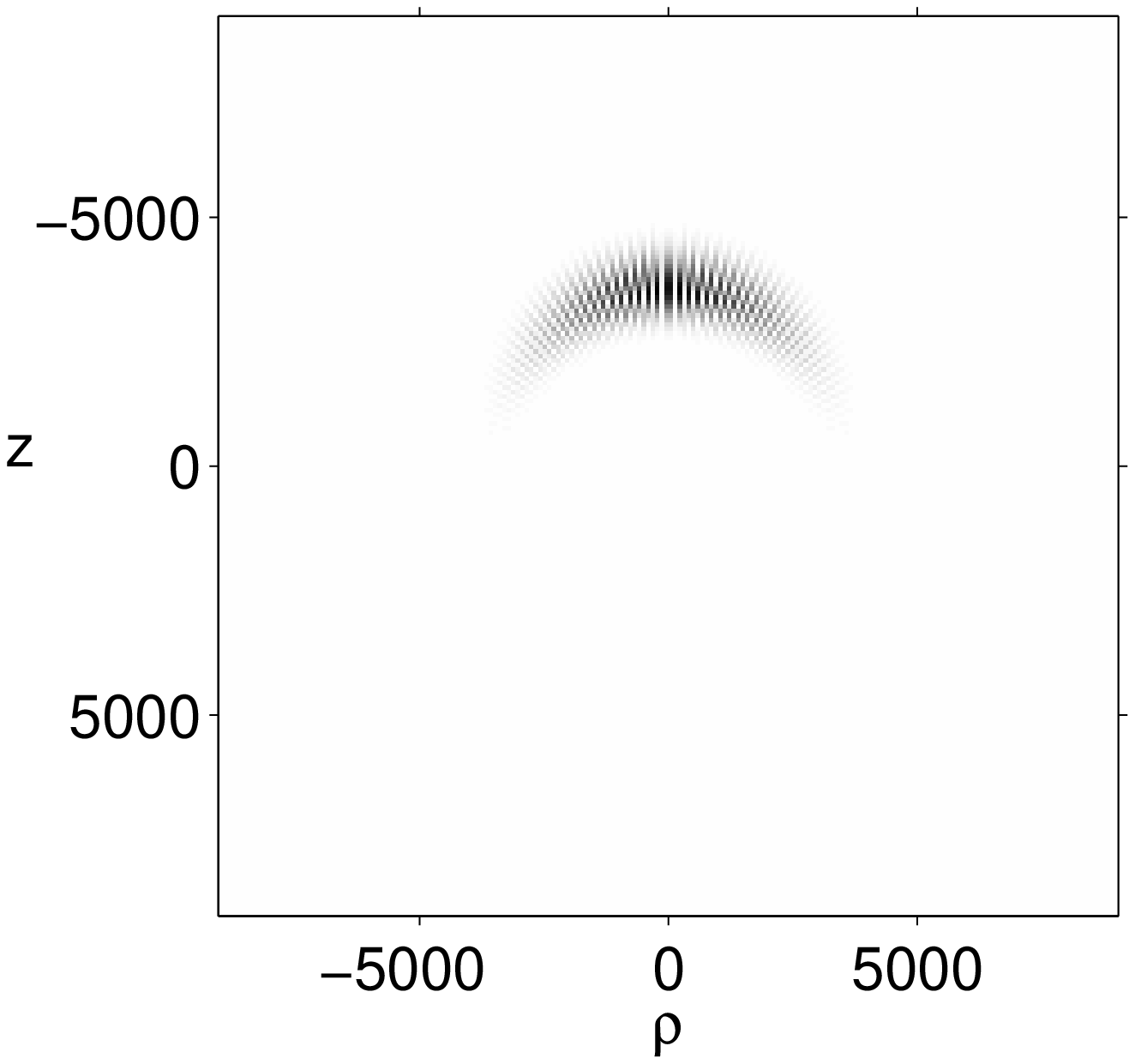}
 \includegraphics[width=0.4\textwidth, clip = true]{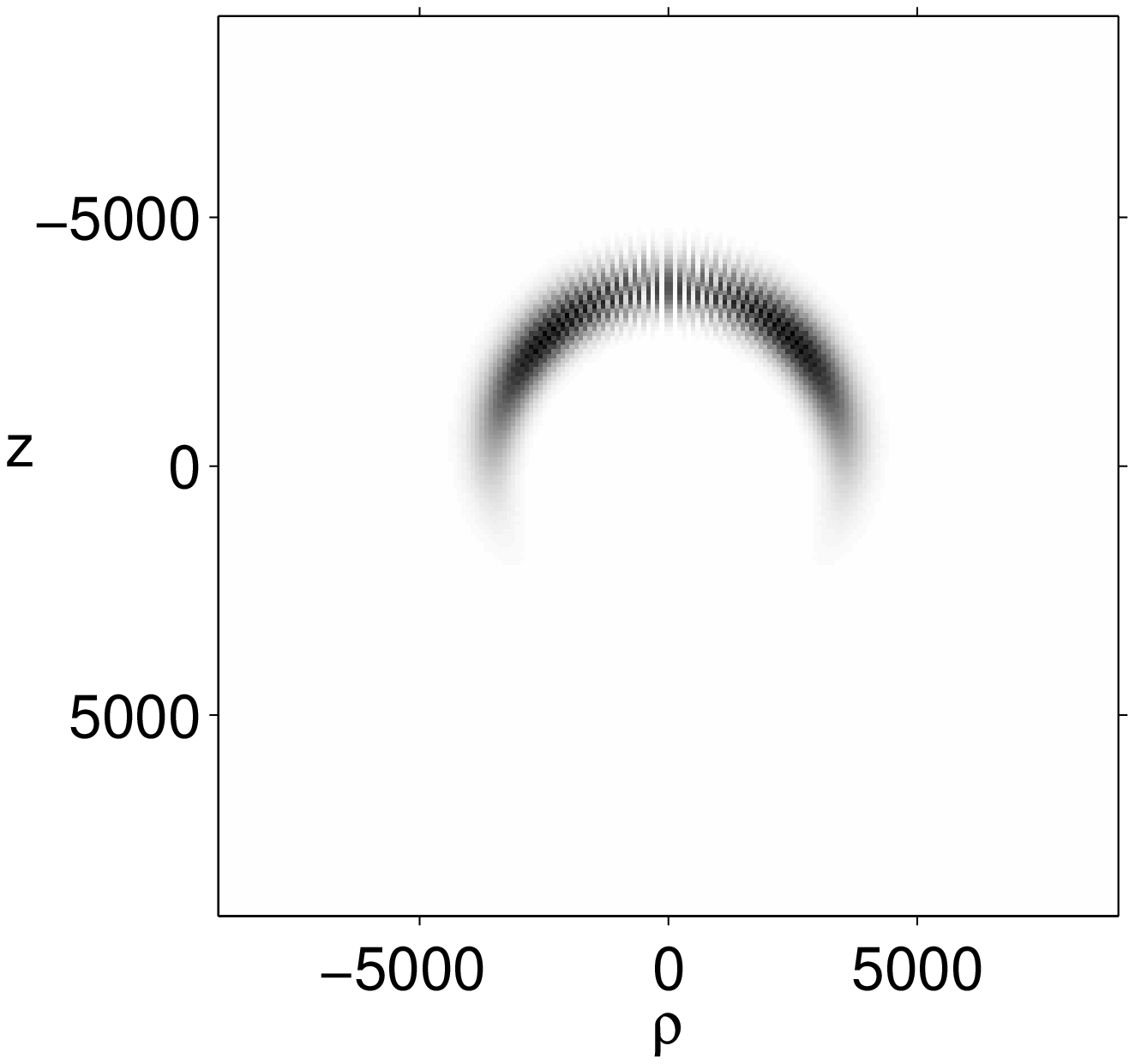}

 \includegraphics[width=0.4\textwidth, clip = true]{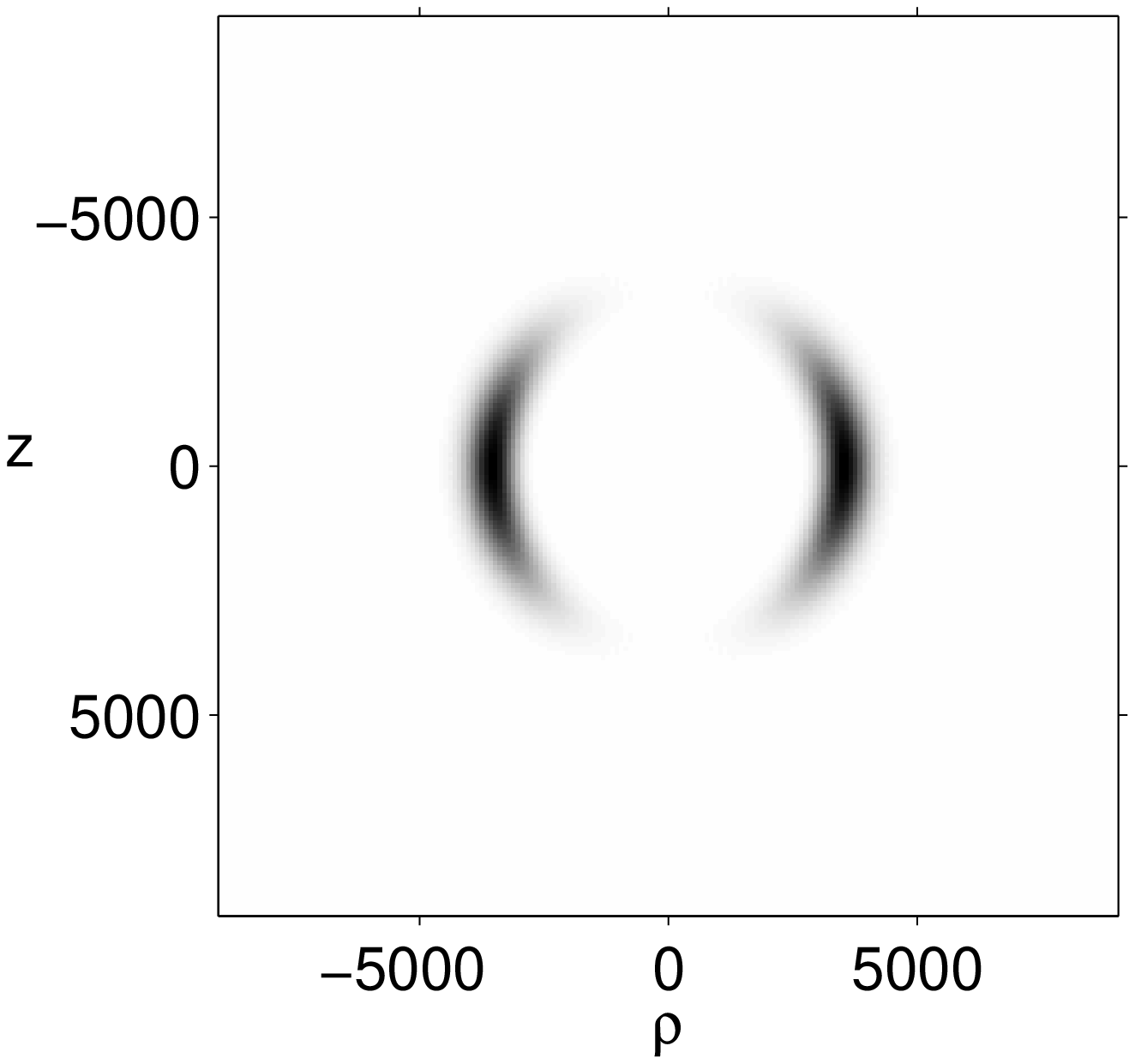}
 \includegraphics[width=0.4\textwidth, clip = true]{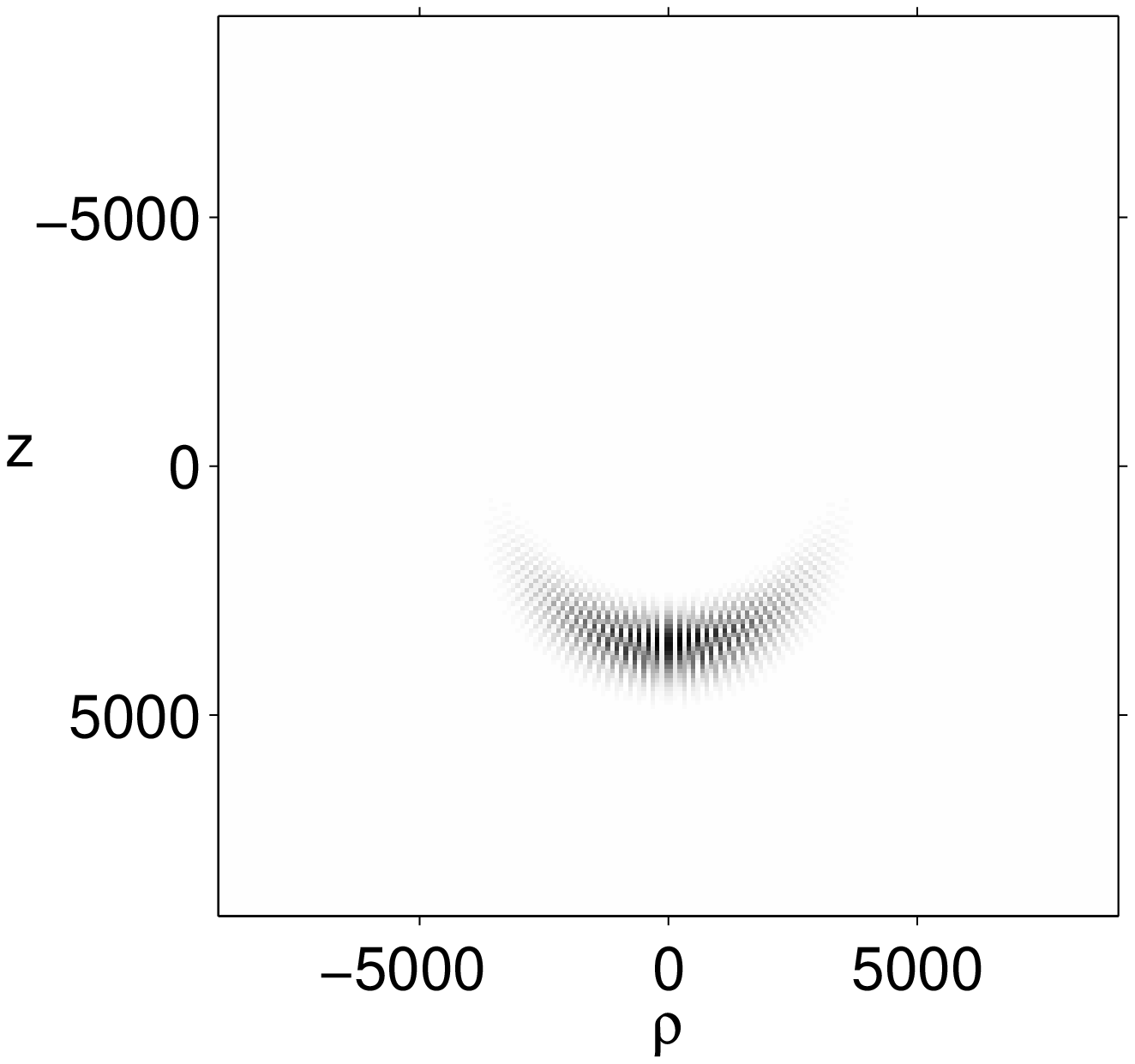}

\end{center}
\caption{The wavepacket (the highest single Floquet state of the $n=60$ manifold in
Fig.~\protect\ref{levdyn2}) obtained from diagonalization of
 the Floquet Hamiltonian for scaled microwave field amplitude $F_0=0.015$ and
 no static field $F_{s,0}=0.0$ and different phases of the
 microwave field (top-left $\varphi=0$, top-right $\varphi=\pi/4$,
 bottom-left $\varphi=\pi/2$ and bottom-right $\varphi=\pi$).
  The wavepacket is a torus
 pulsating on a sphere between the north and the south poles --
 the figure shows a cut along an arbitrary plane containing the $z$ axis
 multiplied by $\rho$ to simulate the density in cylindrical coordinates. The
 $\rho$ on the horizontal axis is either $x$ or $y$ or any other direction
 in the $xy$ plane. The scales are in atomic units.}
\label{floquet_circular}
\end{figure}

\begin{figure}
  \begin{center}
 \includegraphics[width=0.4\textwidth, clip = true]{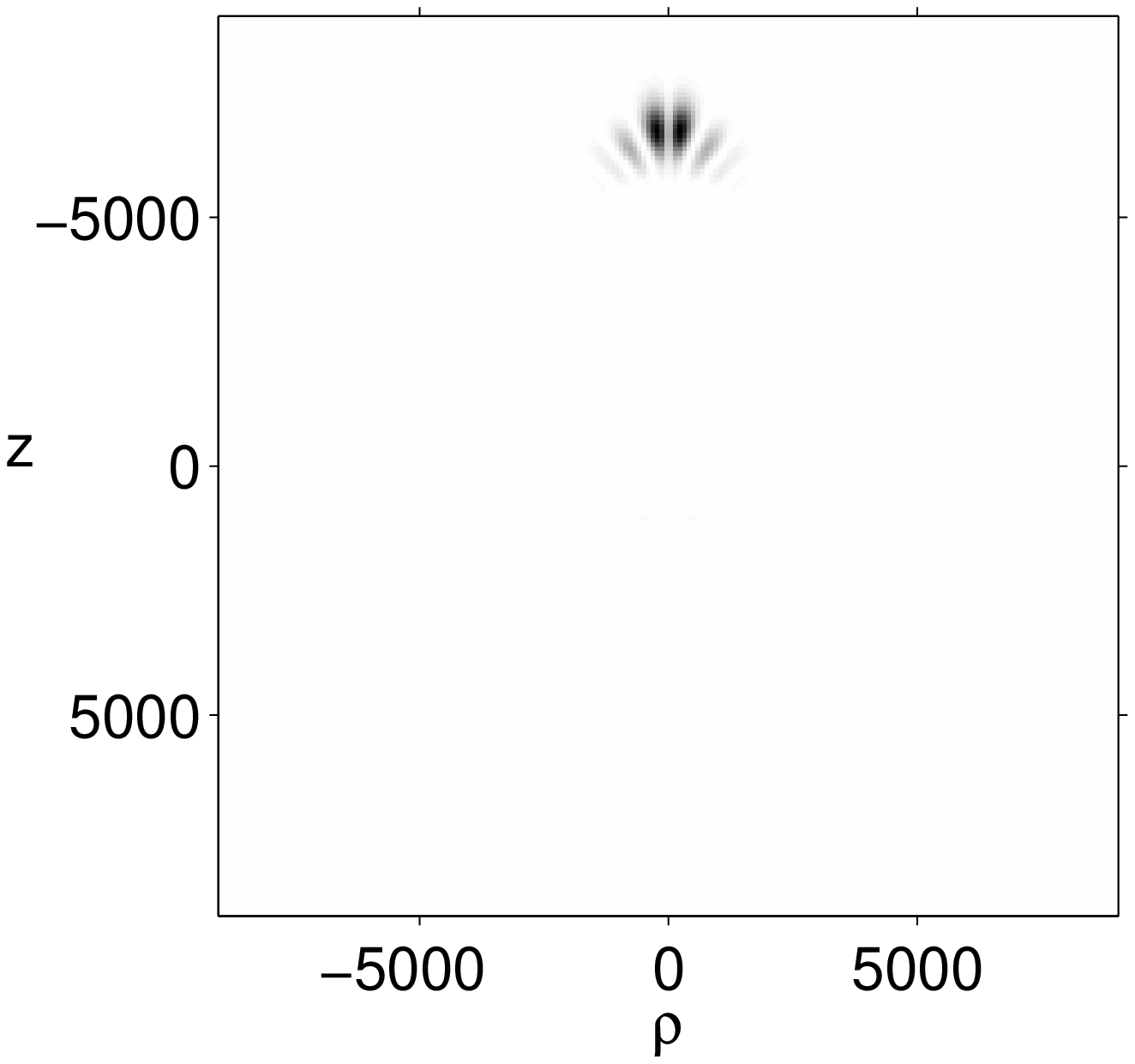}
 \includegraphics[width=0.4\textwidth, clip = true]{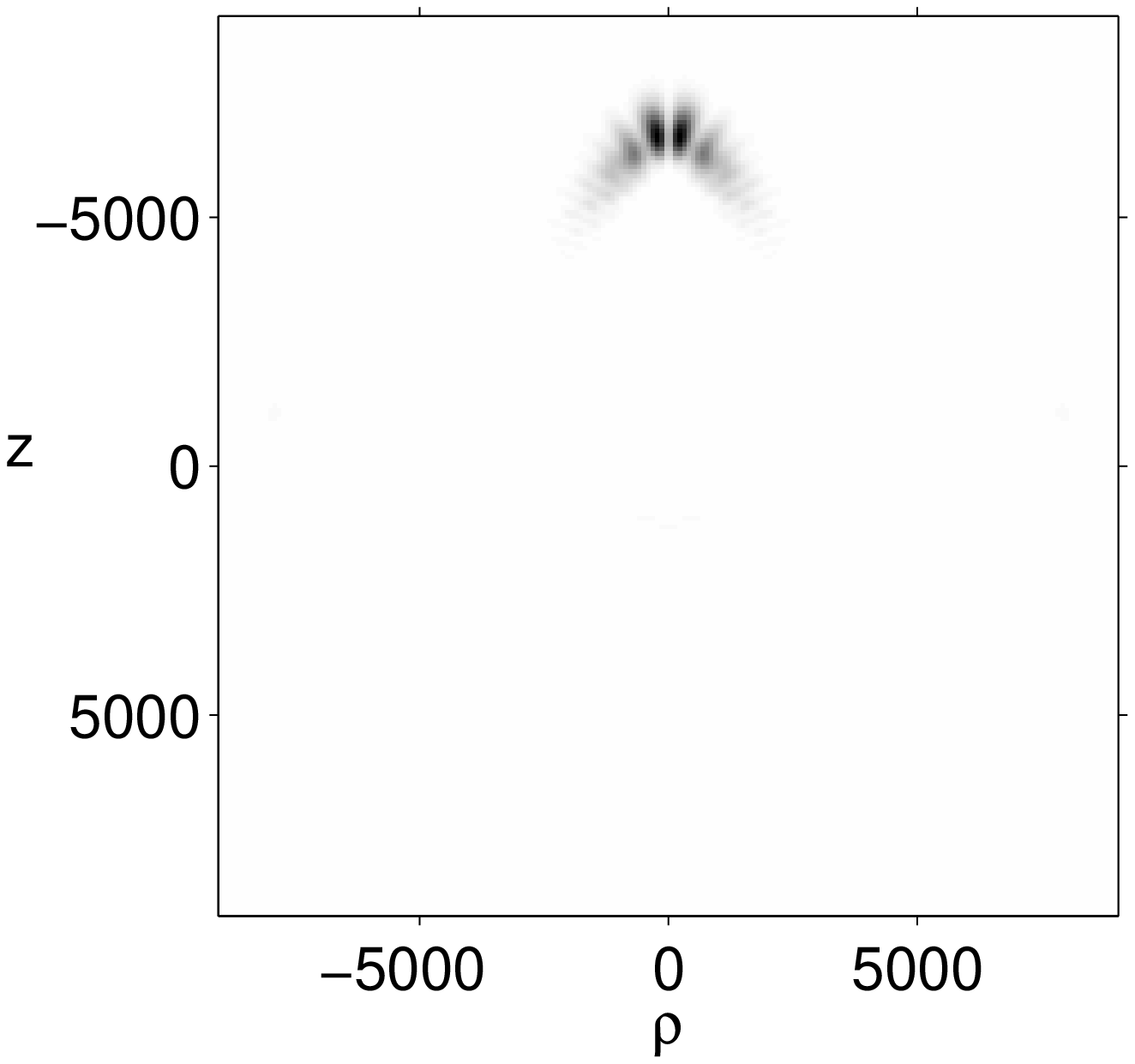}
  
 \includegraphics[width=0.4\textwidth, clip = true]{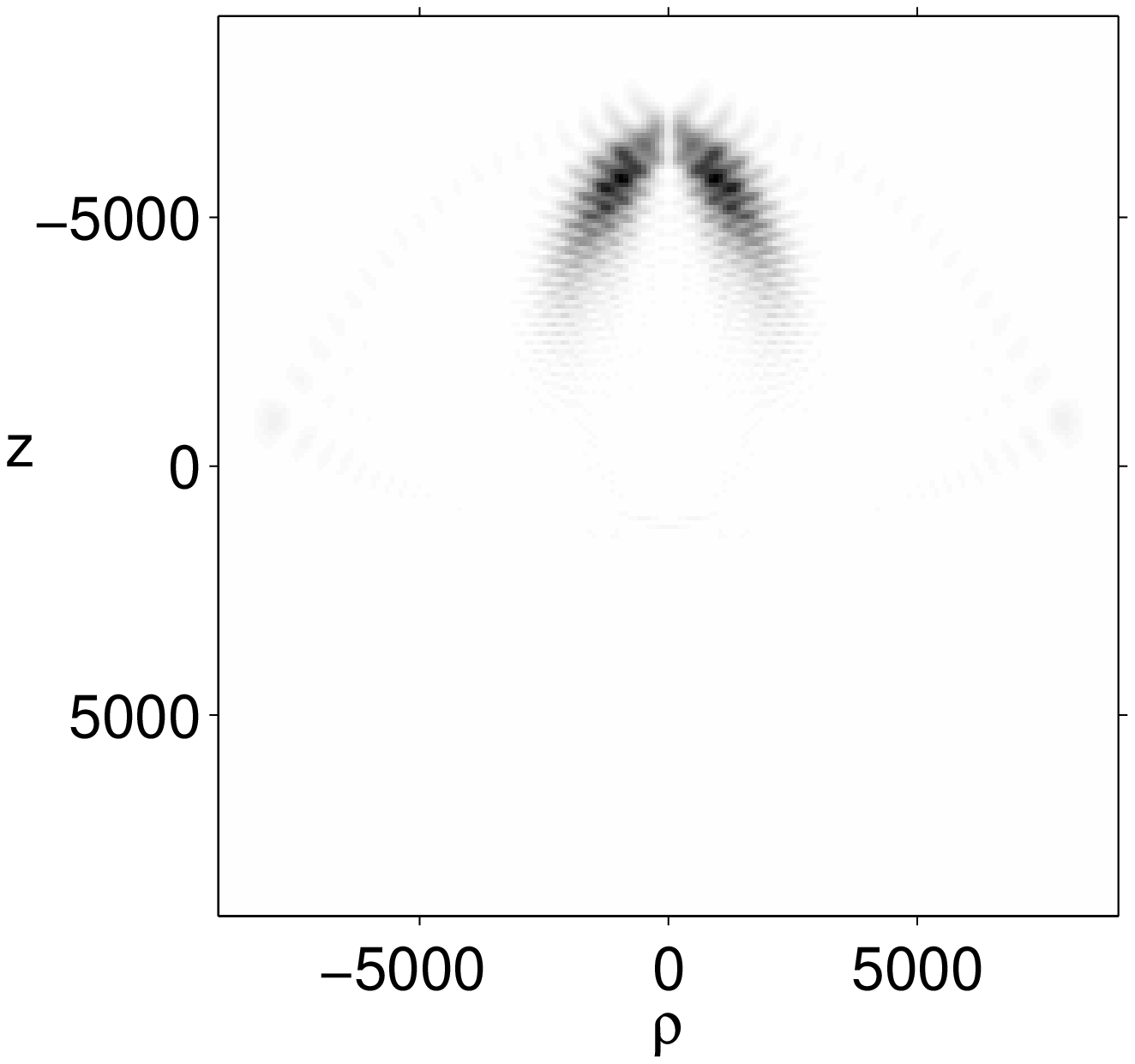}
 \includegraphics[width=0.4\textwidth, clip = true]{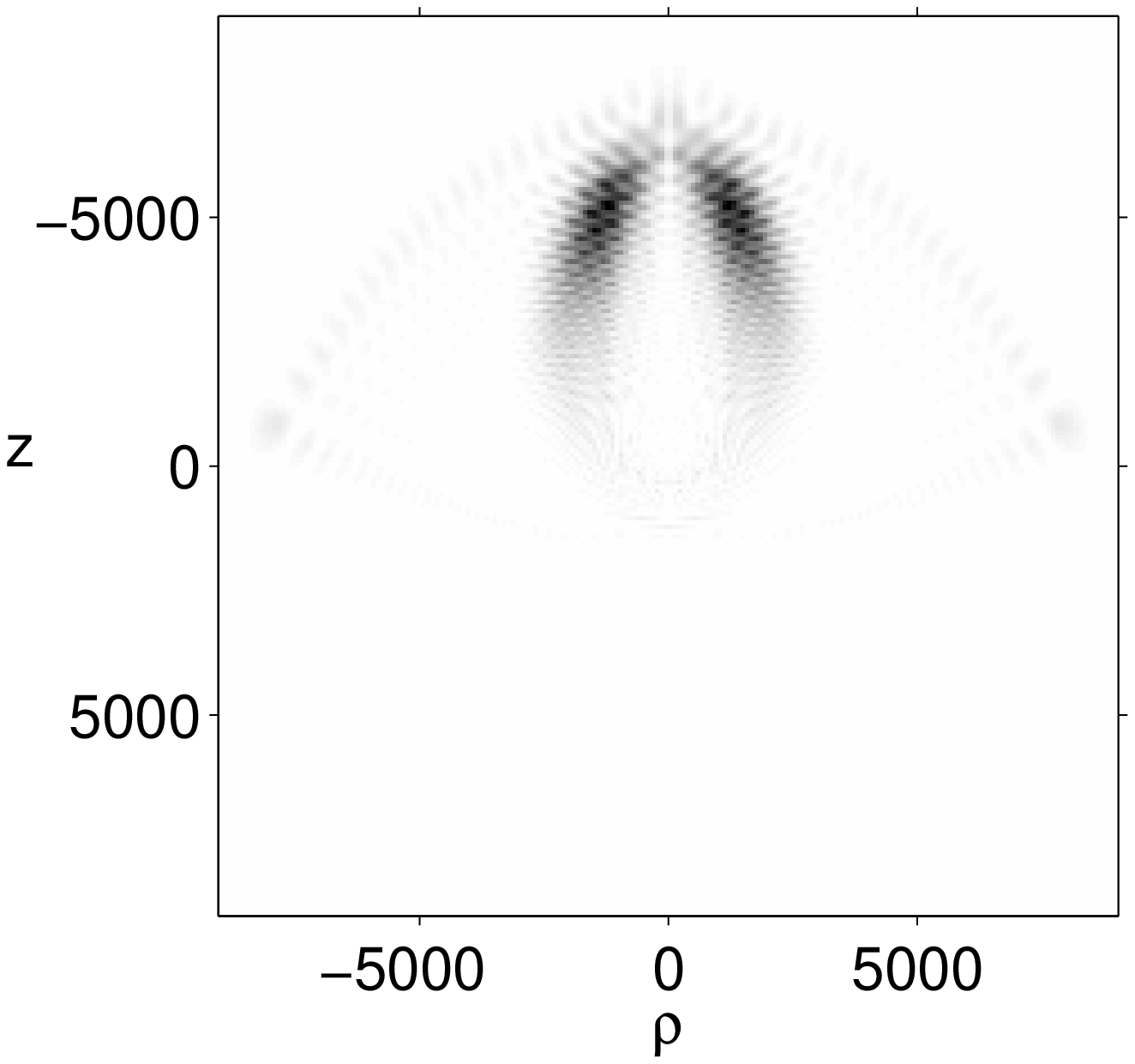}

 \includegraphics[width=0.4\textwidth, clip = true]{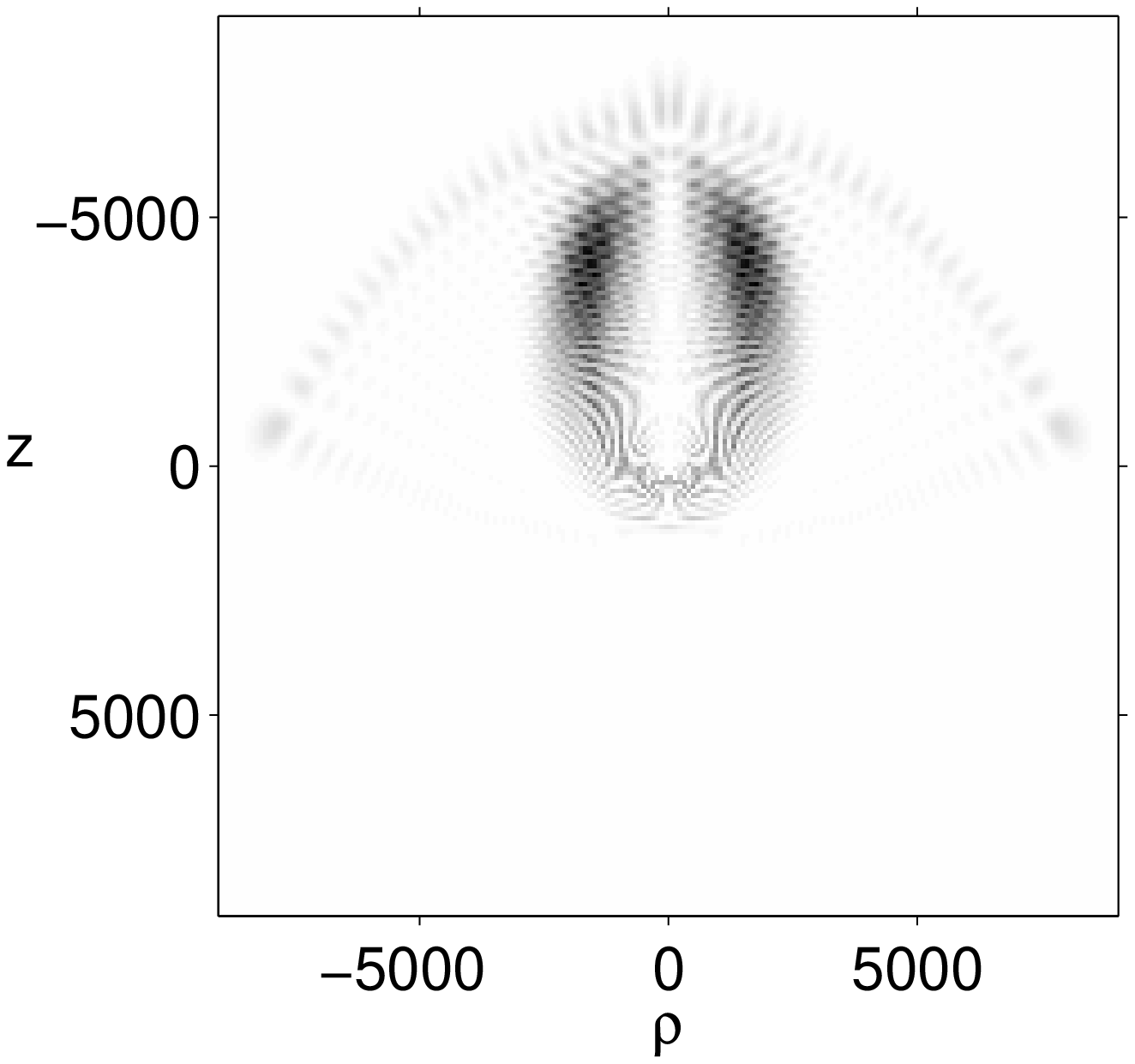}
 \includegraphics[width=0.4\textwidth, clip = true]{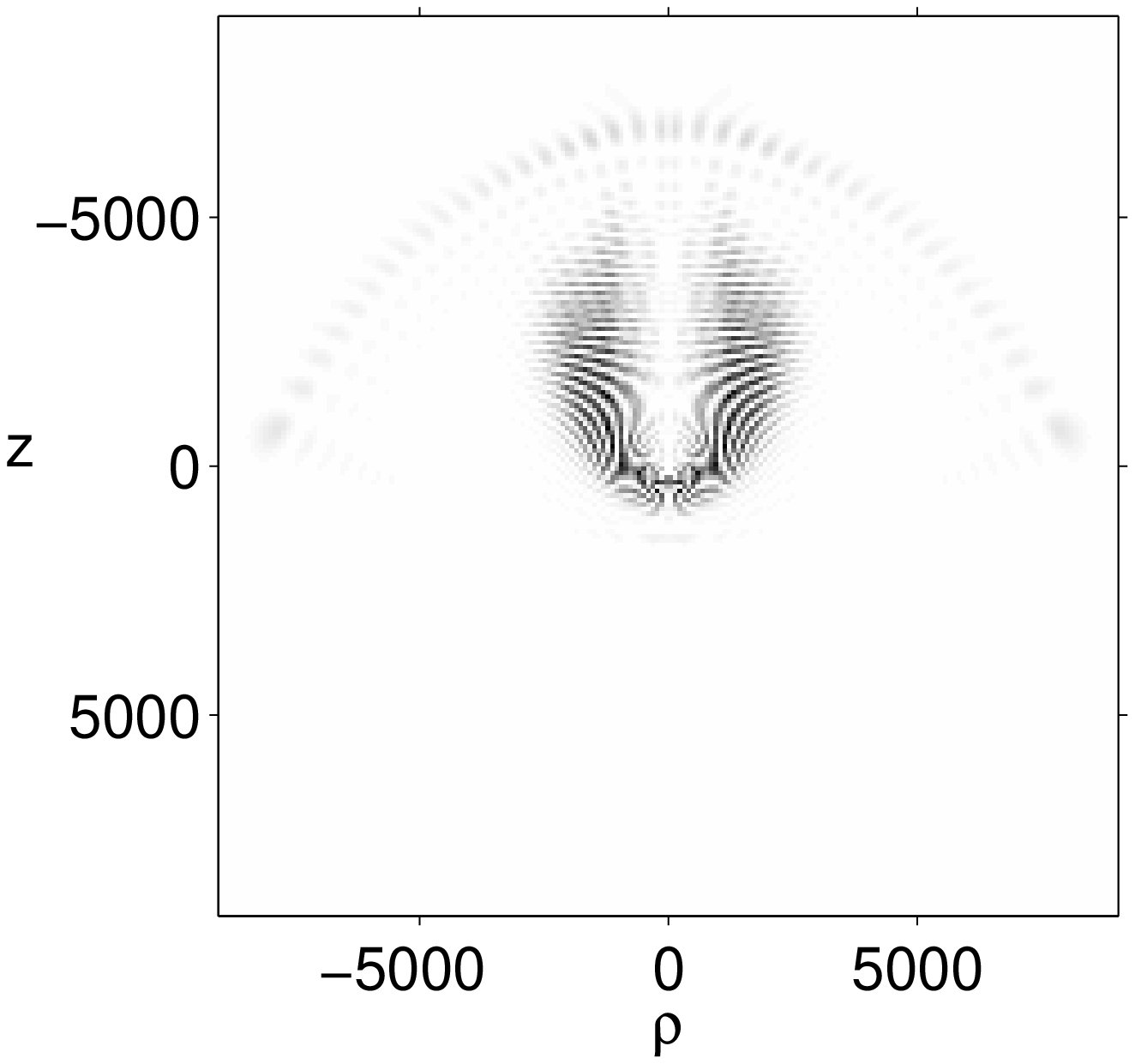}
\end{center}
\caption{The wavepacket (the highest single Floquet state of the $n=60$ manifold in
Fig.~\protect\ref{levdyn2}) obtained from diagonalization of
 the Floquet Hamiltonian for scaled microwave field amplitude $F_0=0.015$ and
 scaled static field $F_{s,0}=0.00255$ and different phases of the
 microwave field 
 (top-left $\varphi=0$, top-right $\varphi=\pi/4$,
 middle-left $\varphi=\pi/2$, middle-right $\varphi=5\pi/8$
 bottom-left $\varphi=3\pi/4$ and bottom-right $\varphi=\pi$).
  For this static field value,
 the nearby avoided crossing leads to a contamination of the wavepacket
  by another state. The wavepacket traces an
 elliptical trajectory. The scales are in atomic units. 
}
\label{floquet_elliptical}		
\end{figure} 

\begin{figure}
  \begin{center} 
 \includegraphics[width=0.4\textwidth, clip = true]{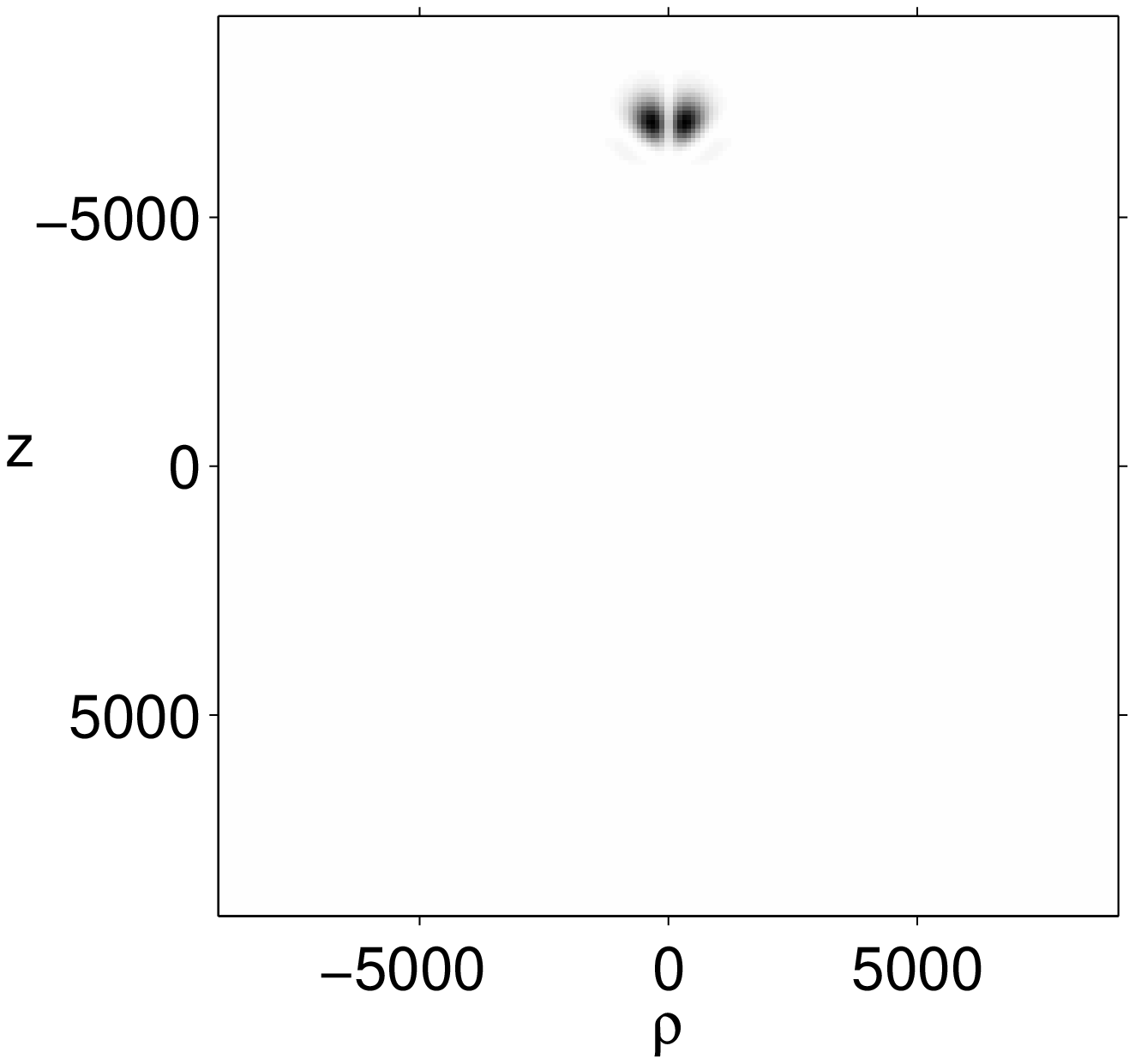}
 \includegraphics[width=0.4\textwidth, clip = true]{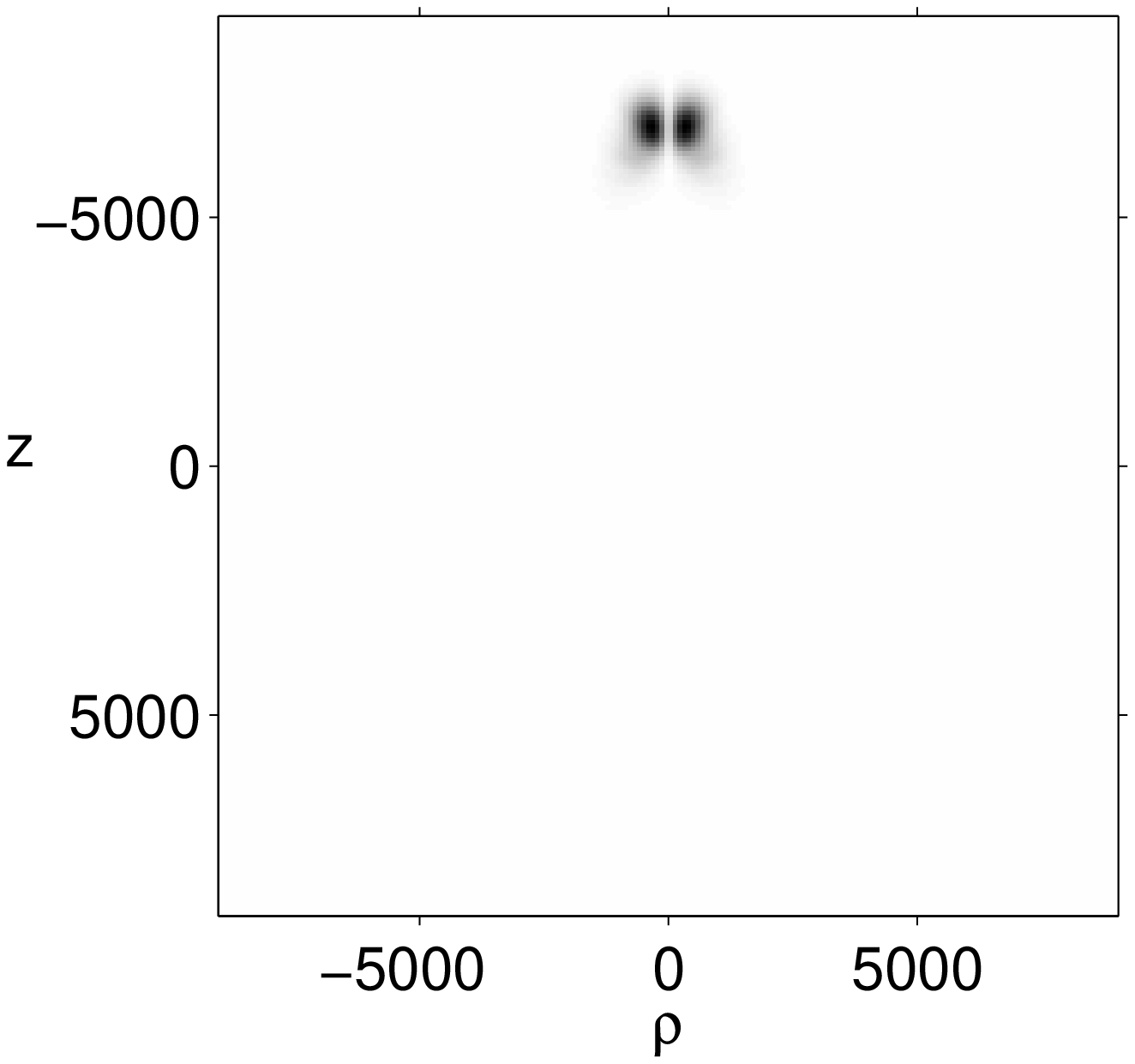}
   
 \includegraphics[width=0.4\textwidth, clip = true]{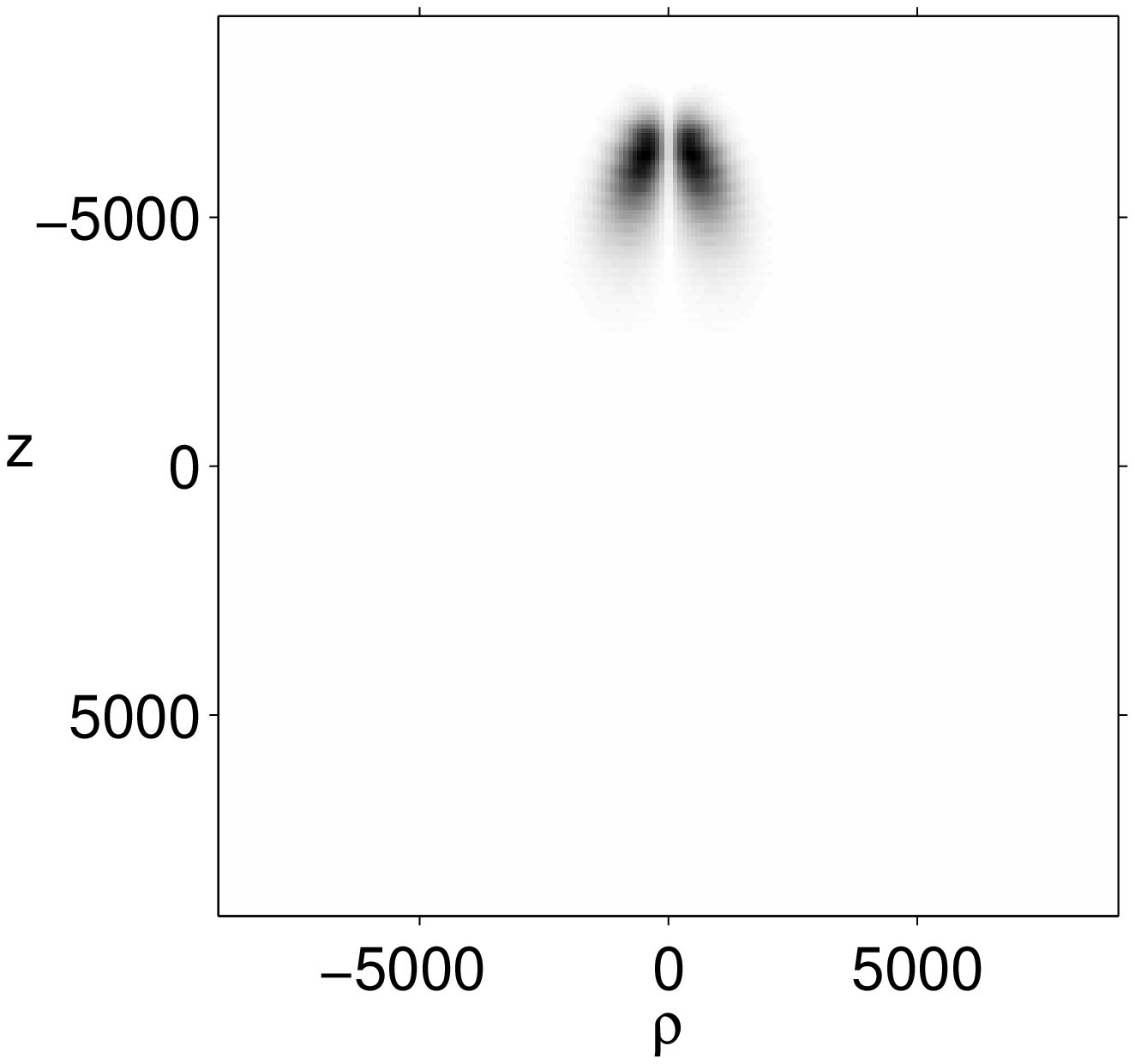}
 \includegraphics[width=0.4\textwidth, clip = true]{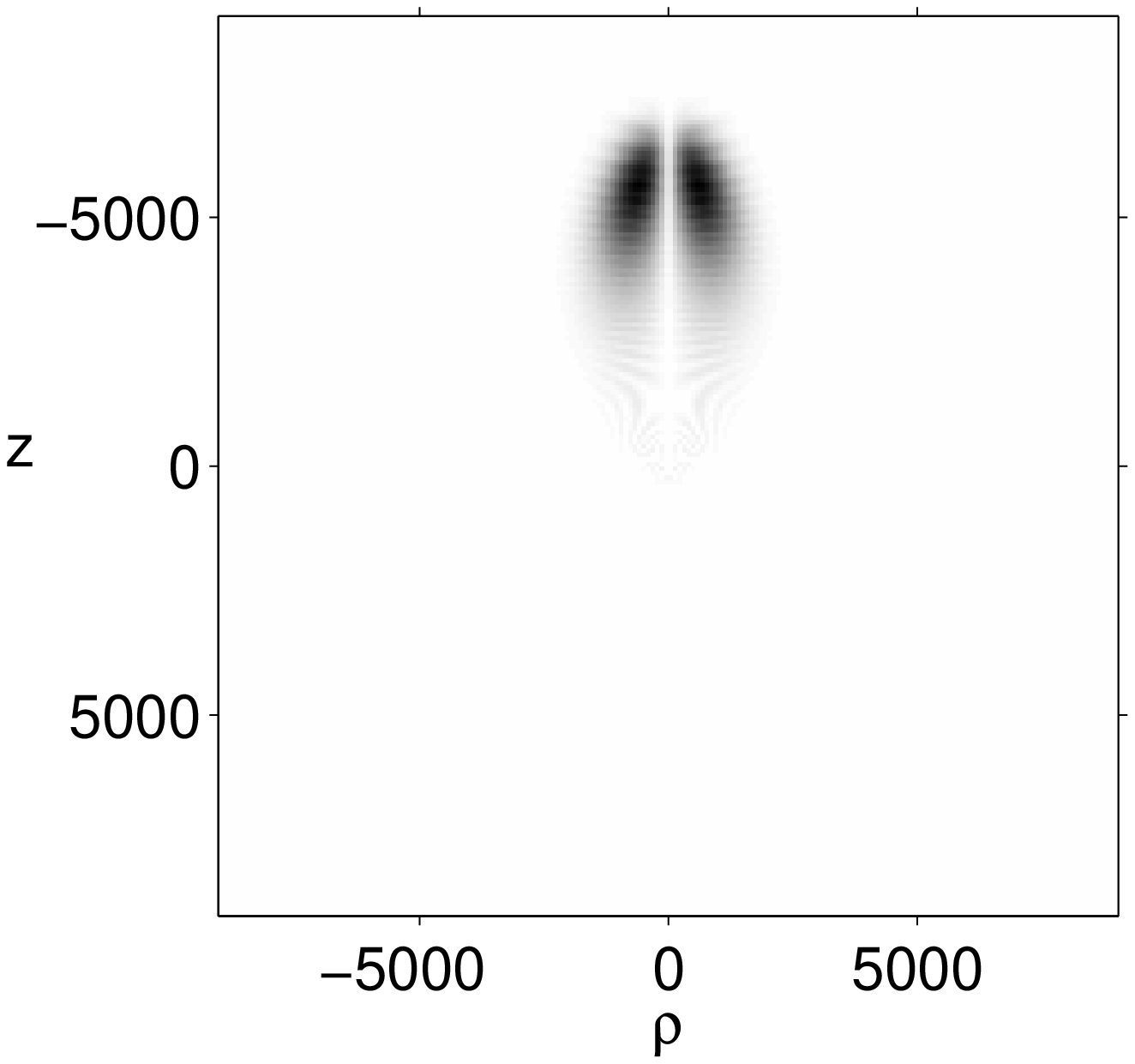}

 \includegraphics[width=0.4\textwidth, clip = true]{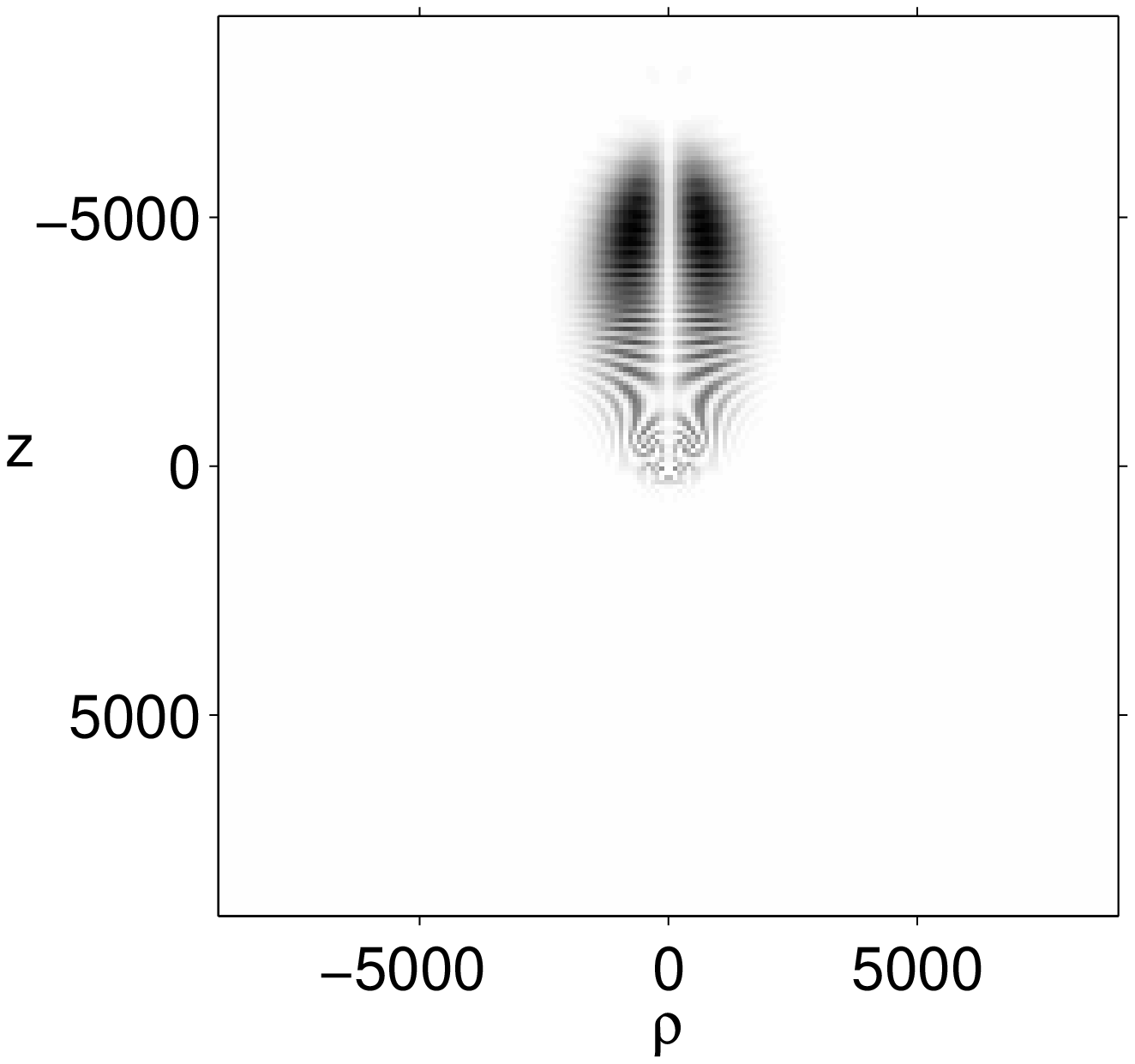}
 \includegraphics[width=0.4\textwidth, clip = true]{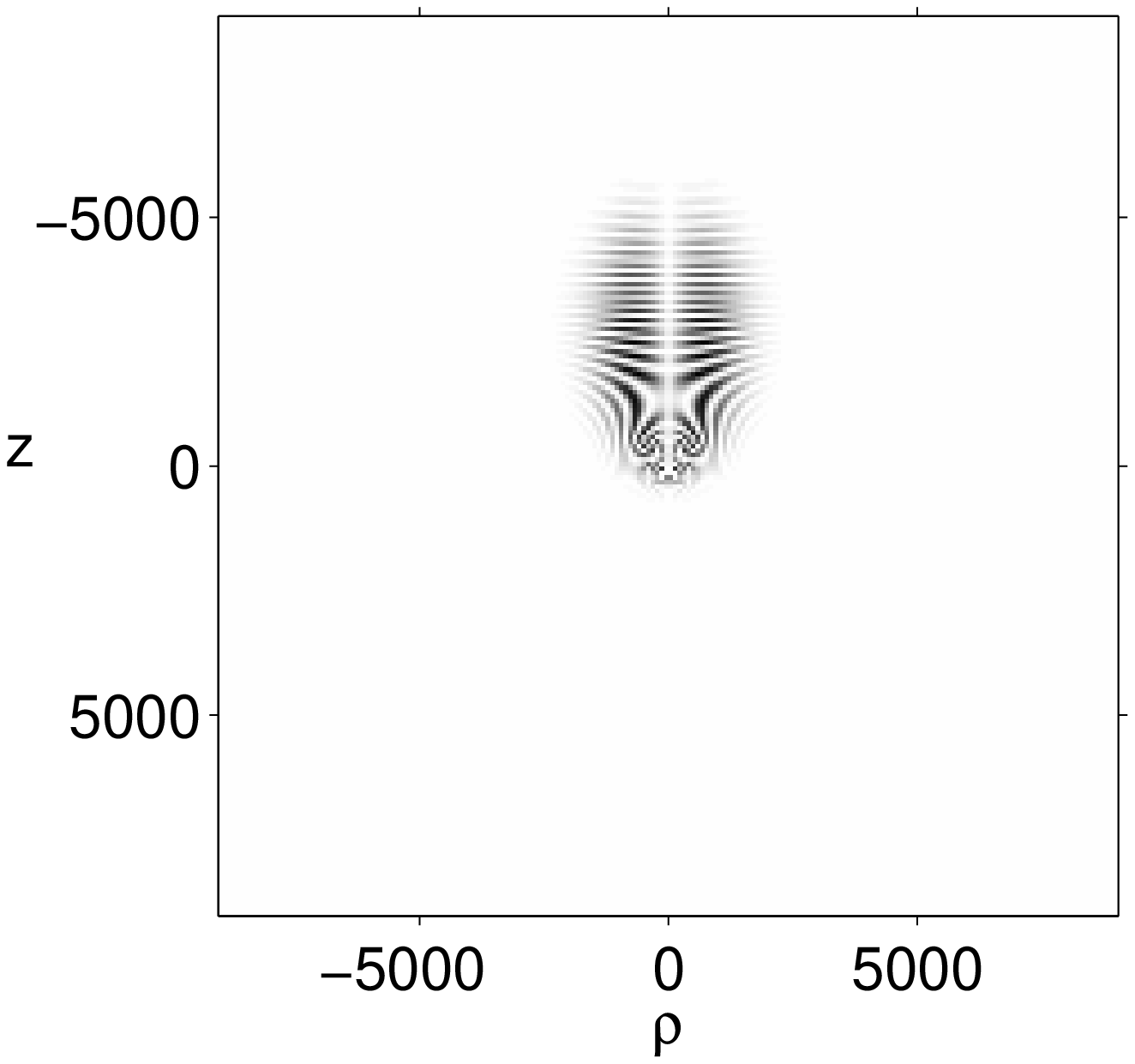}
\end{center}

\caption{The wavepacket (the highest single Floquet state of the $n=60$ manifold in
Fig.~\protect\ref{levdyn2}) obtained from diagonalization of
 the Floquet Hamiltonian for scaled microwave field amplitude $F_0=0.015$ and
 scaled static field $F_{s,0}=0.003$ and different phases of the
 microwave field
 (top-left $\varphi=0$, top-right $\varphi=\pi/4$,
 middle-left $\varphi=\pi/2$, middle-right $\varphi=5\pi/8$
 bottom-left $\varphi=3\pi/4$ and bottom-right $\varphi=\pi$). 
  The wavepacket follows
 a linear trajectory along the field axis. The scales are in atomic units.}
\label{floquet_linear}
\end{figure}

Of special interest is the upper state of the manifold as it has maximum
localization in the resonance island, but also maximum localization
on the classical circular trajectory in the $(\rho, z)$ plane. 
Its temporal evolution over one period of the microwave field,
is shown in Fig.~\ref{floquet_circular}. The plot is obtained from
an exact numerical diagonalization of the Floquet Hamiltonian, following
the techniques described in~\cite{bdg95}. One clearly sees the nondispersive wavepacket
character of this state, which is localized both in the $\rho$ and $z$ directions
at all times, even extremely long (its evolution is periodic
by construction). As expected, the wavepacket propagates along the circular
trajectory with radius given by the Bohr orbit for $n_0=60$, that is roughly
3600 atomic units. In the plot, the wavepacket appears with two components
symmetric with respect to the $z$ axis. Actually, the figure is a cut
of the three-dimensional electronic density by a plane containing the field
axis (multiplied by $\rho$ to simulate the probability 
density in cylindrical coordinates)
 and -- due to the azimuthal symmetry -- must appear symmetric. In the
three-dimensional world, the wavepacket rather appears as a ring propagating
back and forth between the north and south poles of a sphere. The fringes visible
at $t=0$ and $t=\pi/\omega$ are due to interferences between ingoing and
outgoing parts of the wavepacket.

When the static field is turned on, the manifold expands. One clearly sees inside
the manifold the local shrinking of the mean level spacing corresponding to
the hyperbolic fixed points of the transverse dynamics in the
$(L,\psi)$ plane. The upper state of the
manifold is associated with the elliptic (stable) fixed point with 
maximum effective energy in the $(L,\psi)$ plane, which is located
at $\psi=\pi$ (major axis of the Kepler orbit along the field) with 
total angular momentum $L$ decreasing with increasing static field.
Hence, this state is predicted to be a nondispersive wavepacket
with optimum longitudinal localization; it evolves from
a ``circular" wavepacket at $F_s=0$ to a ``linear" wavepacket
above $F_s\simeq 0.2 F$ passing through intermediate
``elliptical wavepacket". 
This is fully confirmed by the exact numerical diagonalization of the  Floquet
Hamiltonian at various static field strengths. We show in
figures ~\ref{floquet_elliptical} and ~\ref{floquet_linear} snapshots
of the electronic densities, which clearly show the evolution
on the classical trajectory as well as the excellent longitudinal
localization of the non-dispersive wavepackets. Note, however, at
$F_{s,0}=0.00255,$ a small contamination of the wavepacket by a 
neighboring state (the field value is intentionally chosen in the
vicinity of a very small avoided crossing) visible by a small
ring of electronic density at 8000 Bohr radii. 

Around $F_{s,0}=0.0028,$ the wavepacket turns into the ``linear" wavepacket.
This transition (actually an inverse pitchfork bifurcation where
the linear trajectory turns from unstable to stable while the
elliptical trajectory coalesces with the linear one and disappears)
is visible in both the semiclassical and the quantum energy spectra 
as a local minimum in the energy level spacing. Above this
bifurcation, the microwave field appears essentially as a perturbation
of the static field, and the whole manifold is approximately composed
of equally spaced levels, like a usual Stark manifold of the hydrogen atom.

\begin{figure}
\begin{center}
\includegraphics[width=11cm]{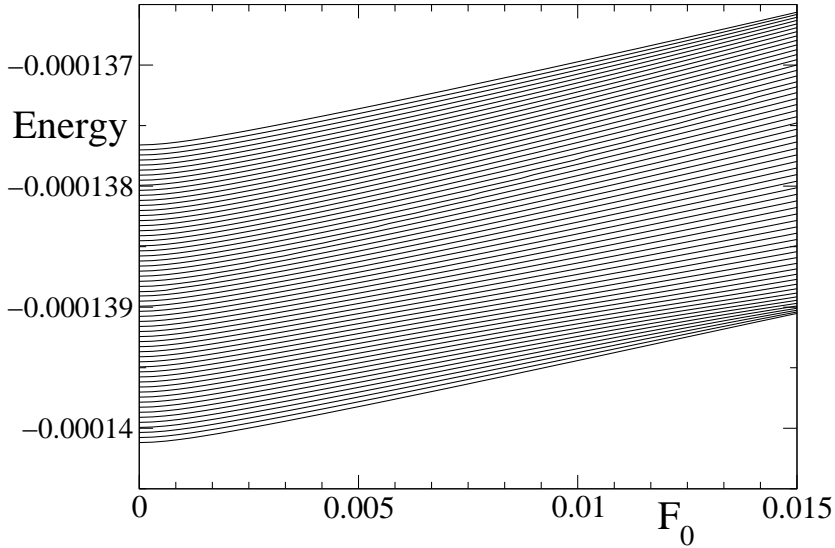}
\end{center}
\caption{Quasienergy levels of the hydrogen atom exposed to parallel static and microwave fields.
They are here plotted for $n_0=60$ -- i.e. microwave frequency 
30.48 GHz -- static field strength $F_{s,0}=Fn_0^4=0.003$ (i.e. 1.2V/cm),
as a function of the scaled microwave field amplitude $F_{0}.$ The  levels of the $n=60$ manifold
are calculated using the semiclassical approximation described in the appendix. The highest level
of the manifold is the non-dispersive electronic wavepacket. For $F_{0}=0,$ it is the extreme
blue shifted Stark state localized along the field axis. As the microwave field is increased,
the electronic density progressively concentrates and builds a non-dispersive wavepacket evolving
periodically along a linear Kepler trajectory.}
\label{levdyn3}
\end{figure}  

In Fig.~\ref{levdyn3}, we show another level dynamics, now at fixed
scaled static field $F_{s,0}=0.003$ and increasing scaled
microwave field. For clarity, only the
semiclassical spectrum is shown, the exact quantum result being
almost indistinguishable. At zero microwave field strength, we have
a pure Stark manifold; at increasing microwave field, one first
sees a quadratic (in $F_0$) increase of the quasienergies,
in accordance with the weak-field limit discussed in the appendix
followed by a linear regime (the strong-field regime discussed in the
appendix). The most important information is that all levels are practically
parallel in the full range, meaning that one passes very smoothly
from a stationary state (at $F_0=0$) to a well localized non-dispersive
wavepacket at $F_0=0.015.$ This smoothness is another illustration
of the robustness of the non-dispersive
wavepackets. Note that, for the highest state of the manifold,
there is no angular evolution of the electronic density when the
microwave field is turned on. Already at $F_0=0$ the extreme
blue shifted Stark state is well localized along the field axis.
By increasing the microwave field, one gains progressive longitudinal
localization along the orbit as the resonance island in the
$(I,\theta)$ plane grows.

\section{Quantum dynamics with slowly changing 
amplitudes of static and microwave fields}

From the preceding section, it is clear that the exact quantum level dynamics
is extremely close to the semiclassical prediction as well as
being very smooth with
tiny avoided crossings only. This implies that the idea of manipulating
the non-dispersive wavepackets by slowly changing the microwave or static
field amplitudes is a realistic one. 

Let us consider the scheme II introduced above. The first step is the
direct optical excitation of a extreme blue shifted Stark state in the absence of
microwave field. In the plot of Fig.~\ref{levdyn3}, this is the highest state
of the manifold. As its wavefunction is elongated along the field axis
and has a significant value close to the nucleus, optical excitation
from a low lying state is possible with high efficiency. 
Increasing the microwave field value is tantamount to adiabatically 
following the highest state of the manifold from the left to the right
of Fig.~\ref{levdyn3}. As the level dynamics is extremely smooth, a very efficient
adiabatic transfer is likely to be possible. In order to test this hypothesis,
we performed a numerical resolution of the time-dependent Schr\"odinger
equation 
in the presence of a static time-independent field ($F_{0,s}=0.003$)
and a microwave
field with slowly increasing amplitude. We chose
the following shape for the microwave field turn-on:
\be
F_0(t) = F_0^{\mathrm{max}} \sin^2 \frac{\pi t}{2 T_{1}}
\label{sach}
\ee
with $F_0^{\mathrm{max}}=0.015$ and $T_{1}=600$ microwave periods.
The choice of the precise value of the switching time $T_1$ is by no means
critical.

\begin{figure}
\begin{center}
 \includegraphics[width=0.4\textwidth, clip = true]{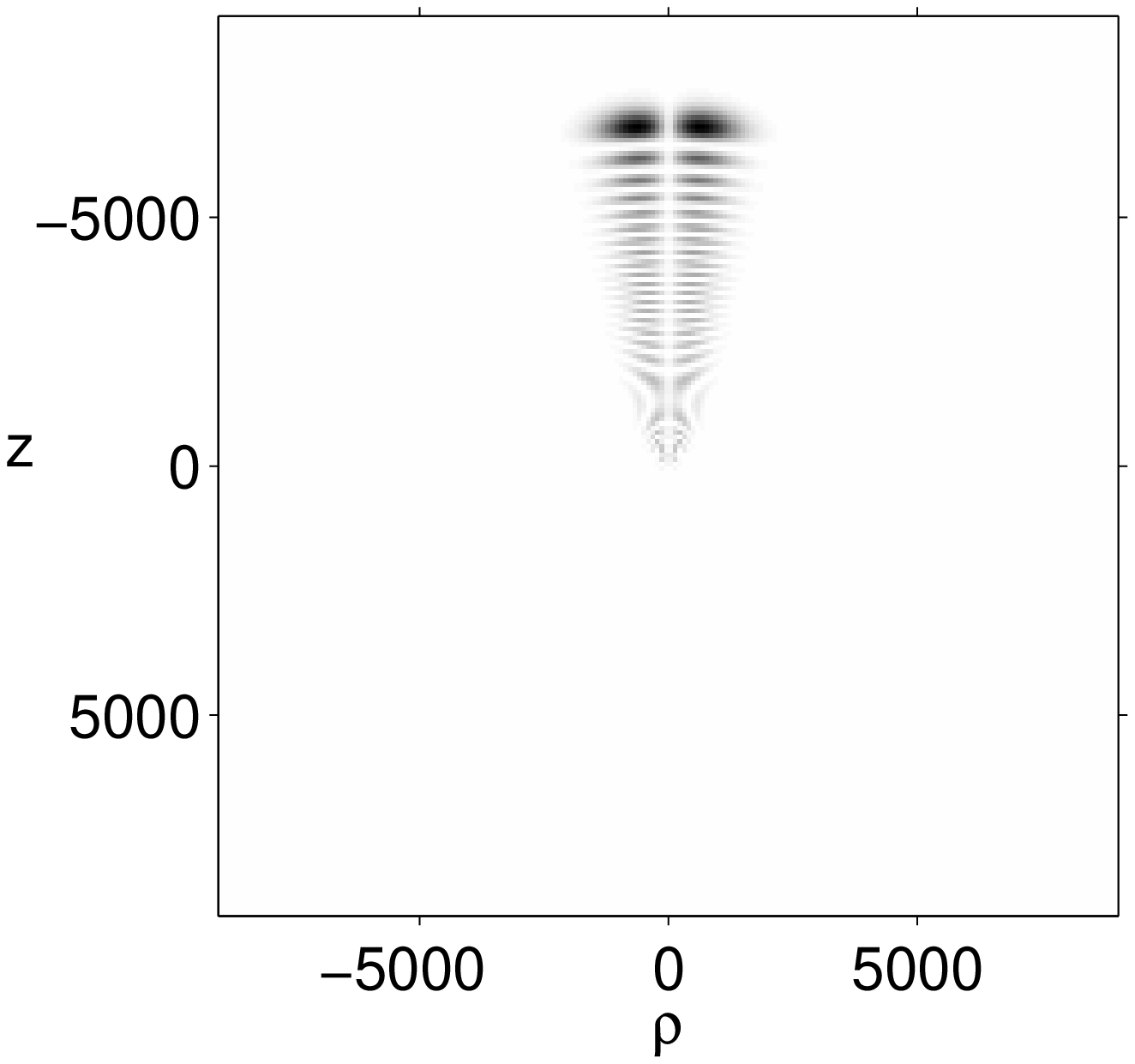}
 \includegraphics[width=0.4\textwidth, clip = true]{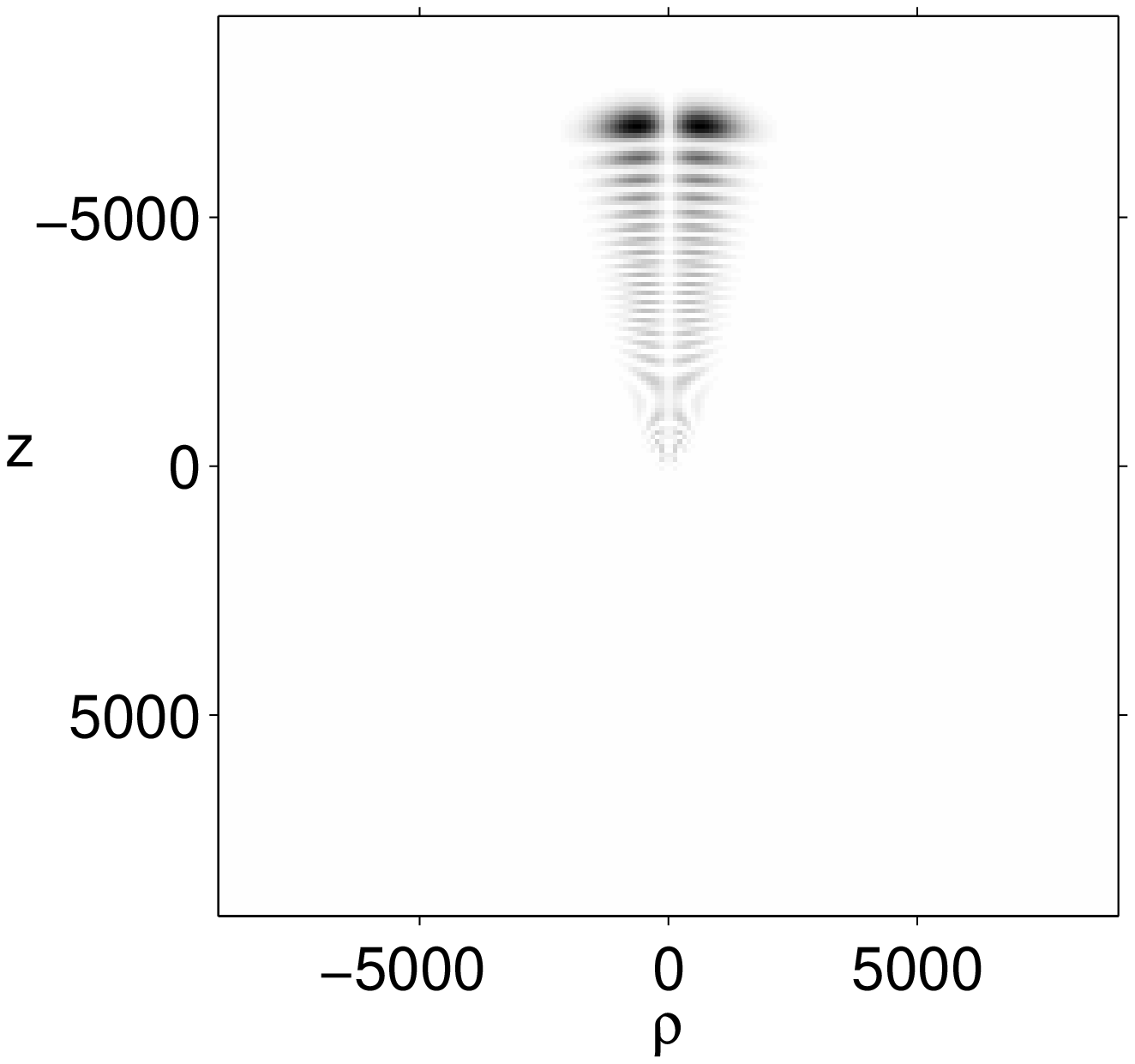}
 
 \includegraphics[width=0.4\textwidth, clip = true]{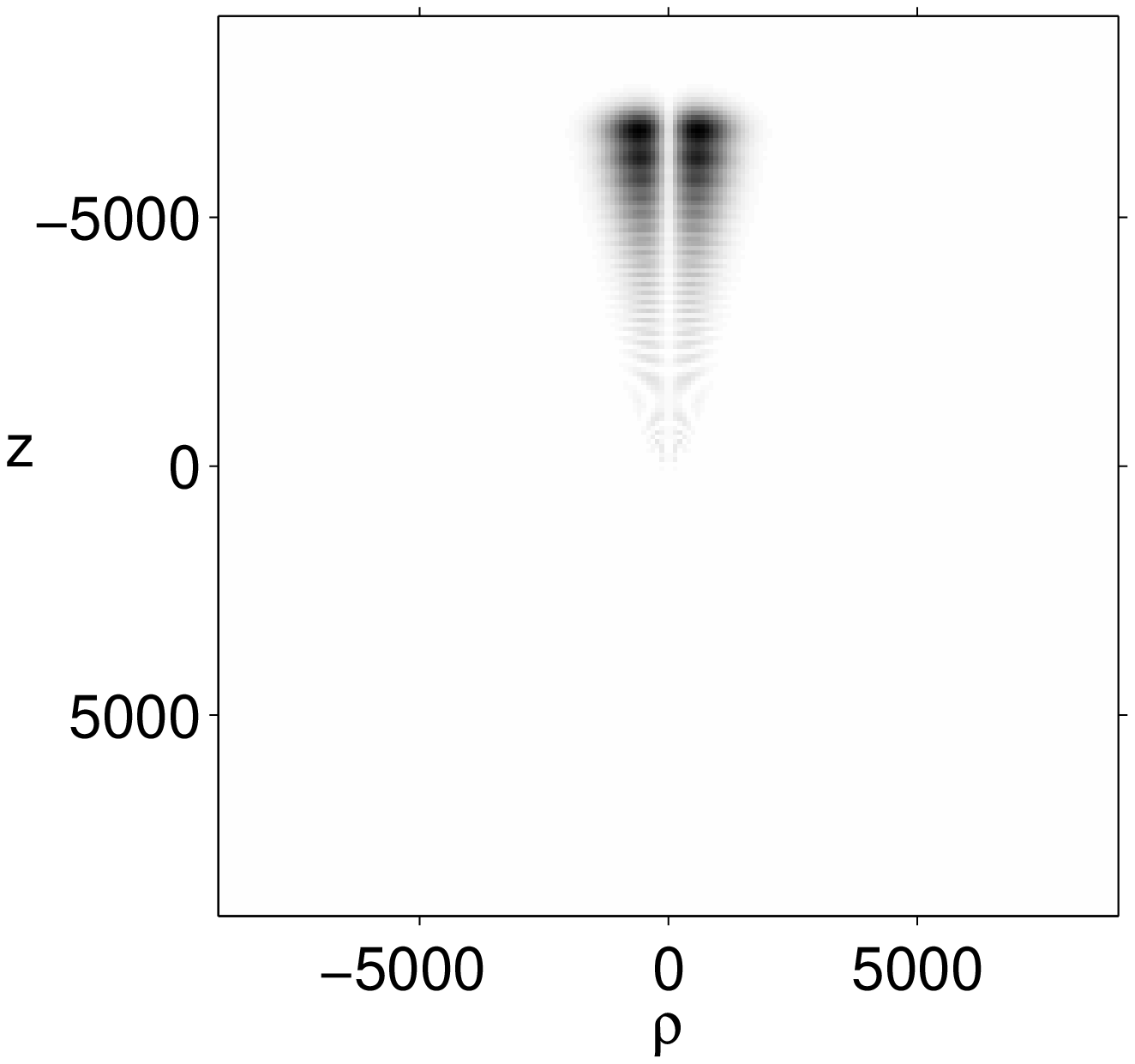}
 \includegraphics[width=0.4\textwidth, clip = true]{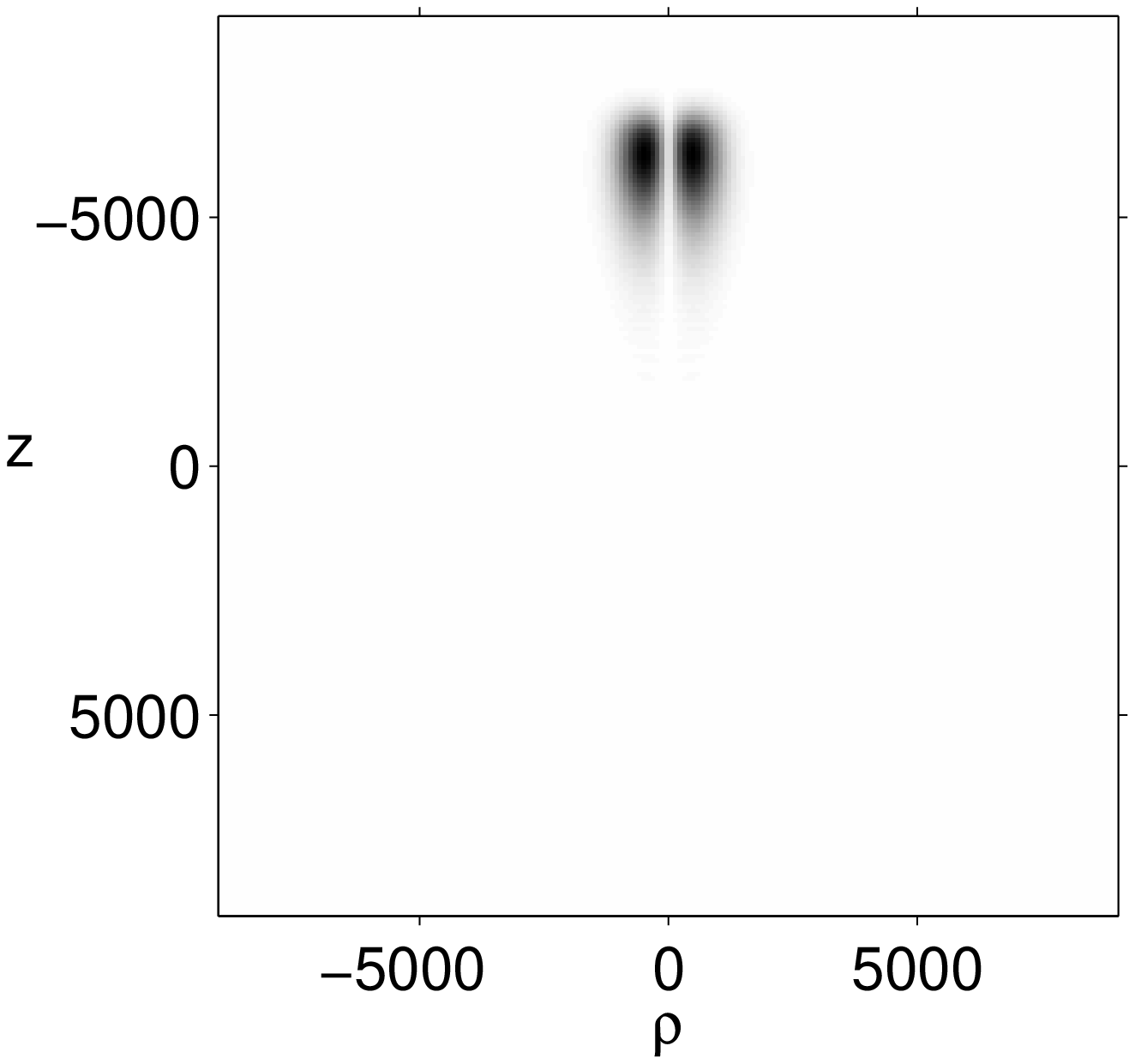}

 \includegraphics[width=0.4\textwidth, clip = true]{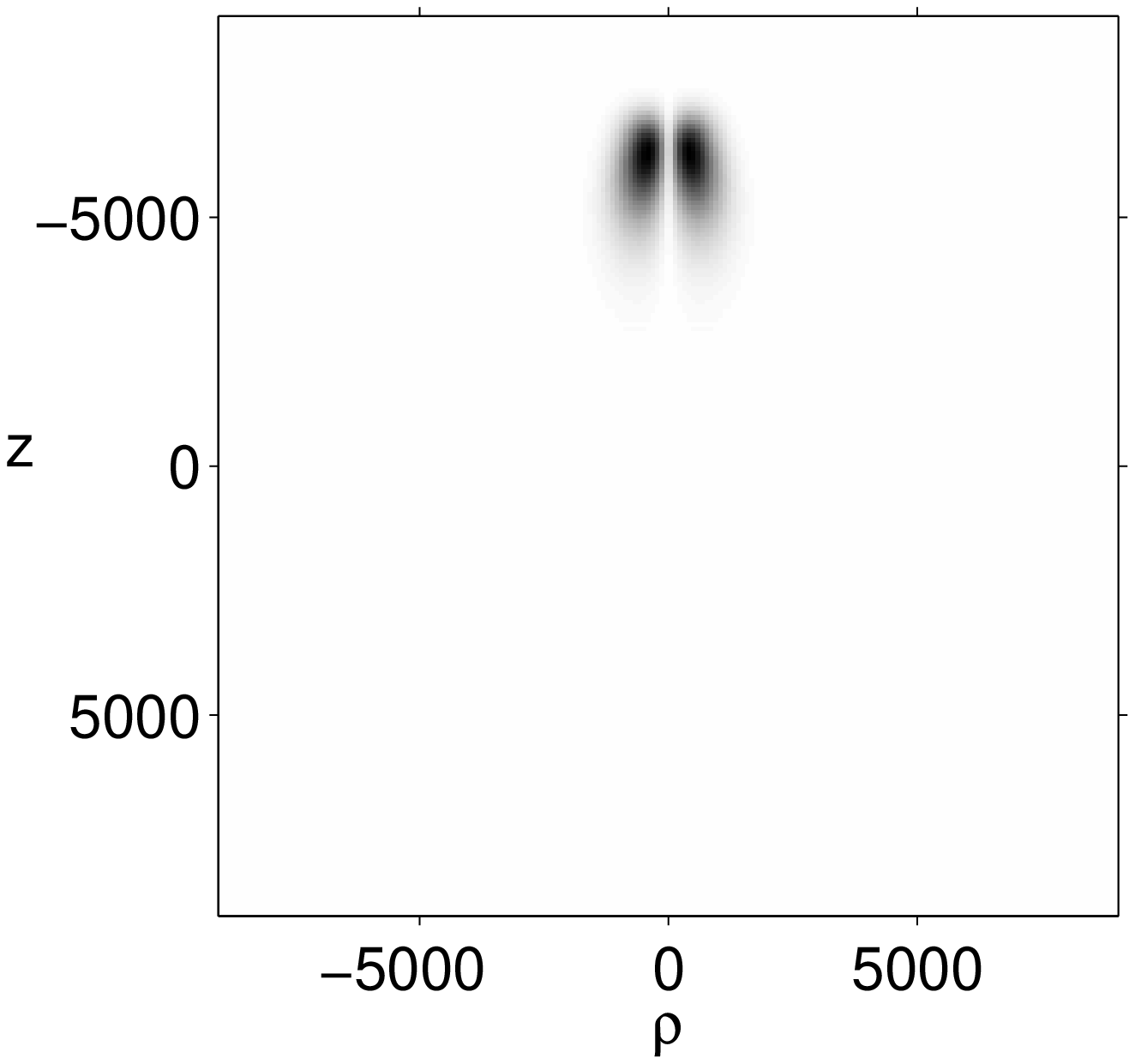}
 \includegraphics[width=0.4\textwidth, clip = true]{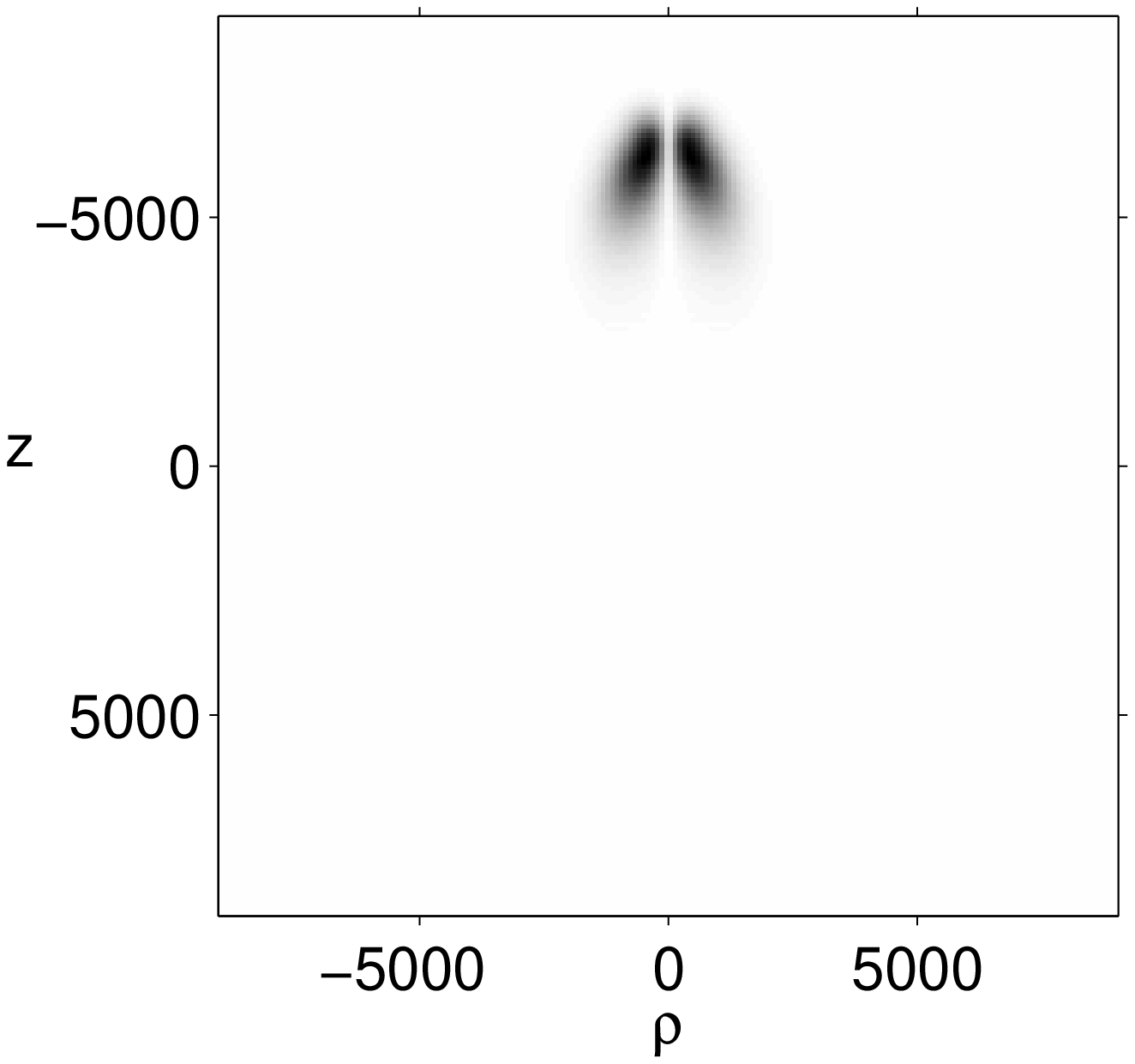}
\end{center}
\caption{Preparation of a non-dispersive wavepacket of the hydrogen atom in
parallel static and microwave fields. The plots show the electronic density obtained
from a numerical resolution of the time-dependent Schr\"odinger
equation at increasing microwave field strengths.
Top left - the initial  state (highest blue shifted state of the
$n_0=60$ manifold); top right
$F_0=0.000367$, middle left $F_0=0.0031$ - already some radial localization
becomes visible; middle right $F_0=0.01$, bottom left $F_0=0.0135$
(these two ones show optimal robust localization) and 
bottom right $F_0=0.015$. The last plot shows a slight tilt indicating the 
vicinity of the separatrix where the linear trajectory becomes unstable.}
\label{td_micro}
\end{figure} 

Snapshots of the electronic density for various values of the
microwave field (at the same phase $\varphi=\pi/2$ of the microwave field) are shown
in Fig.~\ref{td_micro}. They show the progressive localization of the
wavepacket along the classical linear trajectory. The electronic density
of the Floquet state at the same field value is visually not distinguishable
from the one resulting from numerical resolution of the time-dependent Schr\"odinger
equation. For example, the electronic density at $t=T_1,$ thus 
$F_0 = 0.015,$ shown in the bottom-right snapshot in Fig.~\ref{td_micro}
 is almost identical
to the one of the Floquet state in Fig.~\ref{floquet_linear}. 
We calculated the square overlap with the Floquet state as the microwave
field increases and found it to be always of the order of 0.99 or more.

The choice of the switching time and the microwave turn-on function
(\ref{sach}) is not crucial, but
several pitfalls should be avoided:
\begin{itemize}
\item If the switching time is too short, the adiabatic evolution may break down
resulting in the final state being contaminated
 by neighboring states of the same manifold.
This would destroy exact periodicity of the wavepacket and weakly affect its
transverse localization perpendicular to the field.
\item In particular, one should be careful at the very beginning of the pulse
because the wavefunction changes quite rapidly, as manifested by the transition from
a quadratic to a linear dependence of the energy levels with $F_0.$ From that
point of view, it is a good idea to make the microwave amplitude initially
increase like $t^2,$ not like $t$.
\item The switching time should not be too long either, because the
small avoided crossings with states belonging to other manifolds should
be crossed {\em diabatically}. This is not a very severe constraint,
because the avoided crossings are actually small, but switching times
should not be longer than few thousand microwave periods.
\item The non-dispersive wavepackets (as well as other Floquet states) are not exactly 
bound states, but slowly ionize. For the field strengths used here, the ionization
rate is rather small, but increases in the vicinity of the avoided
crossings (see \cite{zdb98}). Less than 1\% of the electronic density is lost
by ionization. However, this process is due to tunneling
and consequently increases very rapidly with the field strength. It may become 
important at higher field values.
\end{itemize}
In practice, our numerical calculations confirm that the excitation of the linear
non-dispersive wavepacket with scheme II can be done with almost 
100\% efficiency in a real
experiment.

\begin{figure}
\begin{center}
\includegraphics[width=6cm,angle=-90]{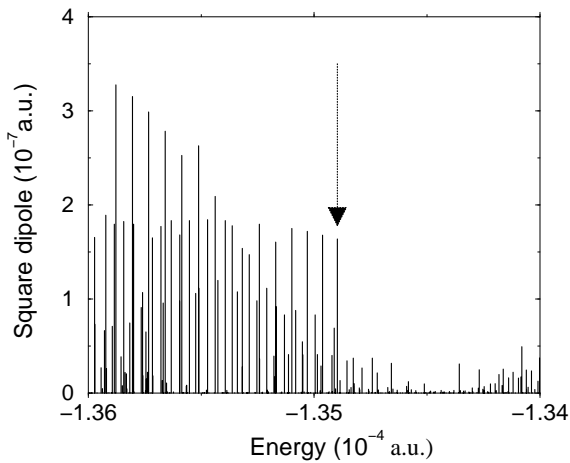}
\end{center}
\caption{Direct optical excitation of the non-dispersive wavepacket, from the ground state
of the atom. The arrow indicates the quasi-energy of the non-dispersive wavepacket propagating
along the field axis. The square dipole of the transition has a rather large value,
which proves the feasibility of direct optical excitation.}
\label{spectrum}
\end{figure} 

Alternatively, scheme I can be used for a direct optical excitation of the
linear
non-dispersive wavepacket, by shining a laser with proper frequency on an atom
in its ground state, {\em in the presence} of the static and microwave fields.
For example, Fig.~\ref{spectrum} shows the excitation probability
(or rather the square dipole matrix element) of the various Floquet states
for $n_0=60$, $F_0=0.02$ and $F_{s,0}=0.006$. The linear
non-dispersive wavepacket is marked with an arrow. It obviously has a significant
excitation probability and we thus believe that the excitation scheme I can also be used.
However, at other field values (such as $F_0=0.015$ and $F_{s,0}=0.003$), it may happen
that several energy levels with significantly higher excitation probabilities
exist at neighboring energies and may hide the state of interest. Also, because
the microwave field considerable increases the effective density of states
which can be optically excited, scheme I requires a better resolution for
selective excitation. This resolution is in the 10-100 MHz range for $n_0=60.$

\begin{figure}
\begin{center}
\includegraphics[width=5cm,angle=-90]{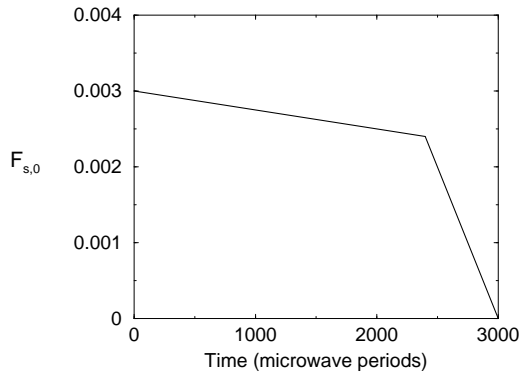}
\end{center}
\caption{By slowly decreasing the (scaled) static field amplitude as shown in this figure,
the circular non-dispersive wavepacket can be prepared with an efficiency greater than 80\%.
The initial decrease must be slow enough to pass the bifurcation
at $F_{s,0}=0.0028$ adiabatically. Once the bifurcation is passed, the decrease
can be made faster.}
\label{piecewise}
\end{figure}

\begin{figure}
\begin{center}
 \includegraphics[width=0.4\textwidth, clip = true]{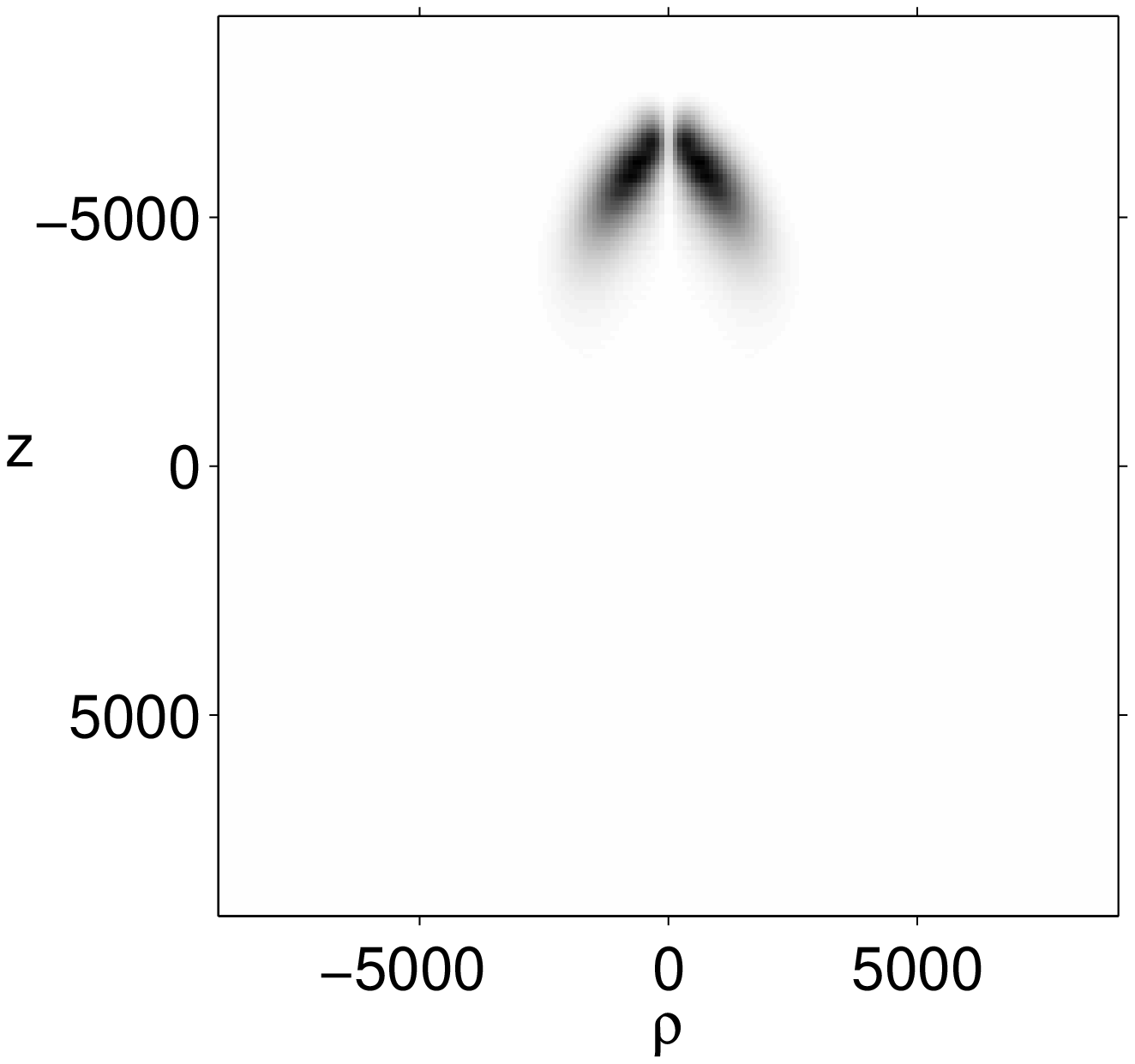}
 \includegraphics[width=0.4\textwidth, clip = true]{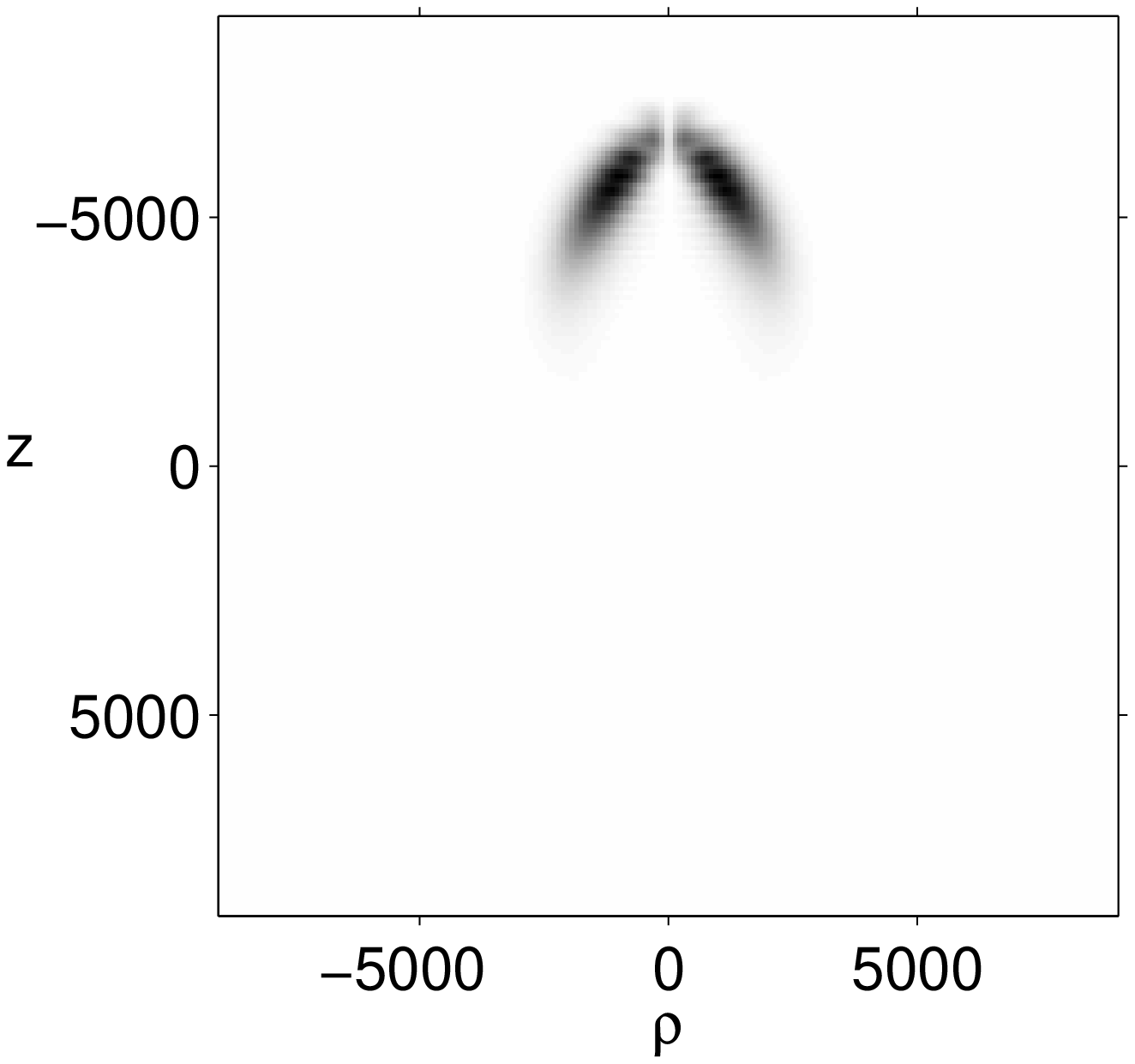}

 \includegraphics[width=0.4\textwidth, clip = true]{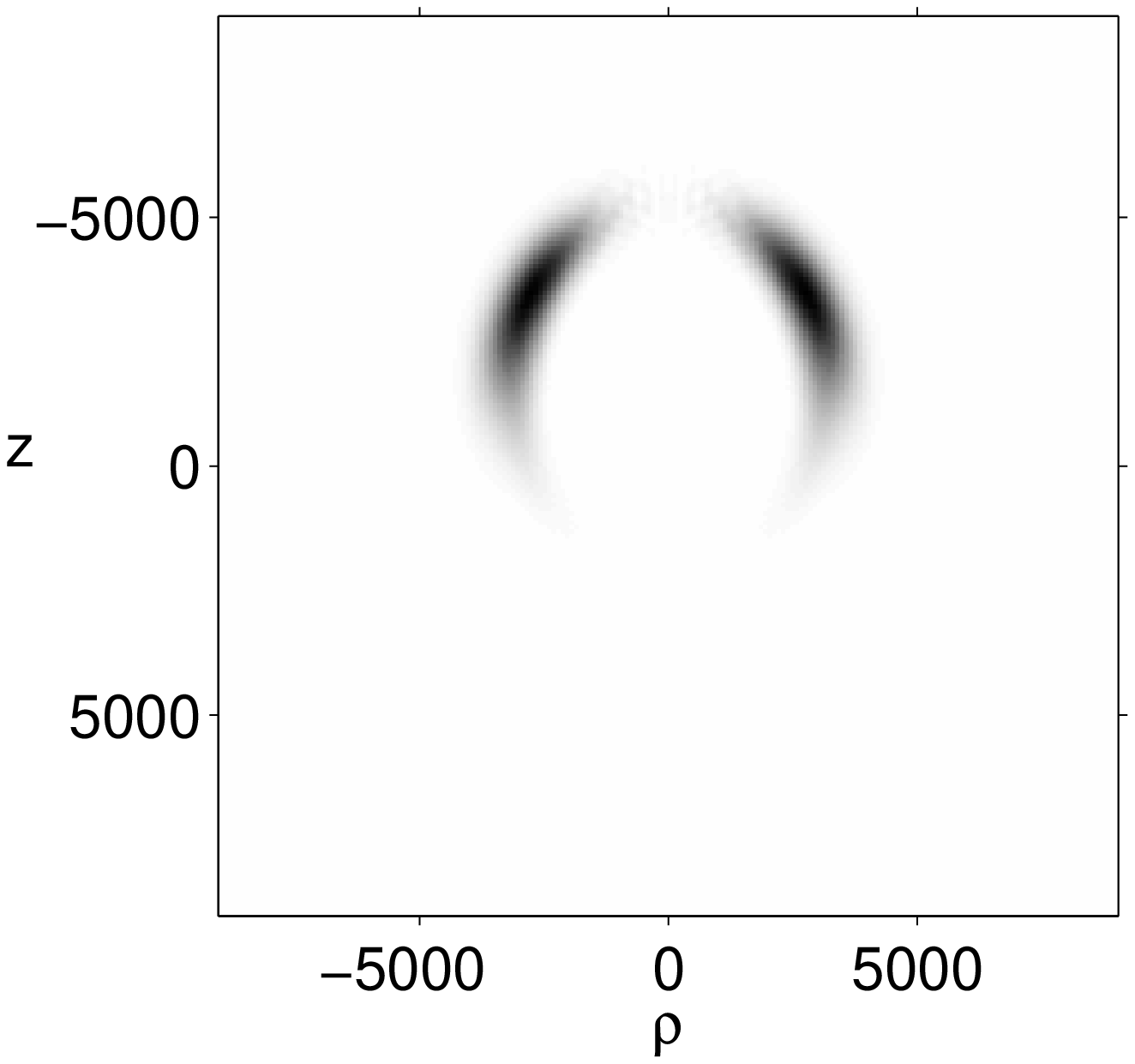}
 \includegraphics[width=0.4\textwidth, clip = true]{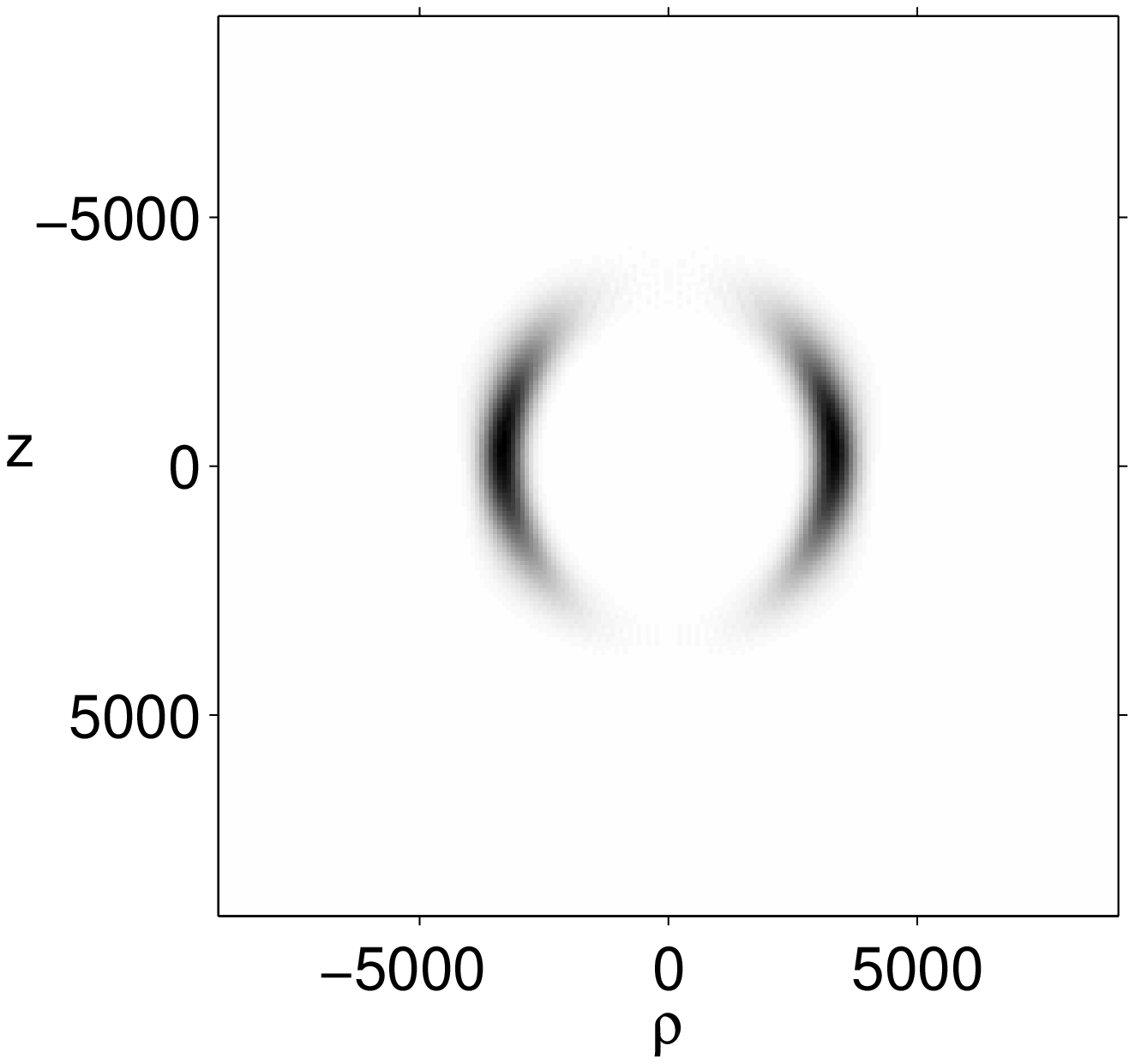}
\end{center}
\caption{Manipulating the wavepacket by controlling the classical trajectory
on which it is localized. Starting from the wavepacket
depicted in the last frame of  figure \protect\ref{td_micro},
the static field is
turned off as described in the text. The plots show the electronic density obtained
from a numerical resolution of the time-dependent Schr\"odinger
equation. Top left $F_{s,0}=0.0027$,
top right $F_{s,0}=0.00255$ (the wavepacket follows an elliptical trajectory),
bottom left $F_{s,0}=0.00144$ (elliptical trajectory with low eccentricity)
and bottom right $F_{s,0}=0$ - the circular trajectory.}
\label{td_static}
\end{figure}

Once the non-dispersive wavepacket on a linear trajectory is created,
the same mixed diabatic/adiabatic transfer can be used in order to transform
the linear wavepacket into an elliptical or a circular non-dispersive
wavepacket. The idea is to evolve from the right side to the
left side of Fig.~\ref{levdyn2} by slowly switching off the static field while
staying in the highest quasienergy level of the $n=60$ manifold. 
The situation is however here a bit more complicated because of the
classical pitchfork bifurcation occurring near $F_{s,0}=0.0028$ and the corresponding
shrinking of level spacing in the quasienergy spectrum. In order to maintain
an adiabatic evolution -- which is essential to transfer angular momentum
to the wavepacket -- the field must evolve rather slowly in this region.
A rough estimate of the maximum velocity at which the static field can
be decreased can be obtained from the minimum size of the level spacing and the
use of the Landau-Zener formula. It turns out that this could lead to
too long switching time and loss of signal either by transfer to other states
at some avoided crossing or by ionization. A solution is to decrease
the static field slowly when crossing the bifurcation
and faster after. For example, we used a piecewise linear function as
shown in Fig.~\ref{piecewise}: slow decrease from $F_{s,0}=0.003$ to
0.0024 in 2400 microwave cycles followed by decrease to zero in 600 periods.
Snapshots of the electronic density at microwave phase $\varphi=\pi/2$ and 
decreasing static field are shown in Fig.~\ref{td_static}. Again, they are
almost identical to the electronic densities of the highest Floquet eigenstate,
proving that the transfer is very efficient. Especially, note that
at $F_{s,0}=0.00255,$ the time-dependent state does not present the extra
electronic density at large distance which is visible in the
Floquet eigenstate, Fig.~\ref{floquet_elliptical}, which proves
that the small avoided crossing with another state is crossed sufficiently
fast (diabatically) to avoid contamination.

Finally, when the static field is completely switched off, the square overlap with
the desired Floquet state -- a circular non-dispersive wavepacket -- is
slightly larger than 0.80. It proves that the scheme II  that  we propose
is efficient although it involves several steps.

\section{Conclusions}

By numerical resolution of the time-dependent Schr\"odinger equation
under realistic conditions and an analysis based on both
semiclassical and exact numerical diagonalization of the Hamiltonian,
we have shown how to prepare efficiently non-dispersive electronic wavepackets
in the hydrogen atom which propagate either on a linear straight trajectory
along the microwave field or on an elliptical trajectory of arbitrary 
eccentricity
(the circular trajectory being the final state). 

Although we concentrated on specific values of the field frequency and amplitude,
the scheme is rather general and could be used for different parameters, with
the following observations:
\begin{itemize}
\item For too low field amplitudes, the resonance island is so small
that no well localized state actually exists. The rule of the thumb
is that only $q$ values, see eq.~(\ref{def_q}), larger than unity
should be used.
\item For too large field values, the states are well localized, but ionize
rather fast. Scaled microwave field amplitudes exceeding 0.04 are dangerous.
\item Even if ionization remains small, it may happen for 
too large field values that the $n_0$ hydrogenic manifold is so large that it 
overlaps with neighboring $n_0\pm 1$ manifolds in the Floquet spectrum. This
creates large avoided crossings which makes the adiabatic transfer impossible.
\item For $n_0=60,$ the microwave frequency is 30 GHz and the total switching time
if of the order of 120 ns. Such switching times should be feasible in a real experiment.
Going to lower $n_0$ values would lead to higher frequency
(and consequently more expensive microwave equipment) and shorter switching
times. Going down to $n_0=30$ is thus rather a bad idea.
\end{itemize}

Excited wavepackets may be detected by employing short half-cycle pulses that 
lead to considerable ionization of the atom \cite{bs99}. 
The ionization signal depends 
on the position of the center of the packet with respect to the nucleus 
at the moment when the pulse is applied (basically the ionization probability
is larger the closer the center is situated with respect to the nucleus
\cite{bs99}). For a discussion of characteristic properties which would
allow for an unambiguous characterization of the non-dispersive wavepackets,
see~\cite{BDZ02}. 

\section{Acknowledgment}

Support of KBN under project 5P03B08821 (KS and JZ) is acknowledged.
The additional support of the bilateral Polonium and PICS programs is
appreciated.
Laboratoire Kastler Brossel de
l'Universit\'e Pierre
et Marie Curie et de l'Ecole Normale Sup\'erieure is
UMR 8552 du CNRS. We thank IDRIS for providing us with CPU time
on a NEC SX5 computer.

\section{Appendix}

In this appendix, we explain the 
semiclassical quantization procedure that allows us
to obtain accurate predictions for the 
quasienergy levels and structures of the corresponding 
eigenstates of the atom in the presence of static and
microwave external fields.

We begin with the effective Hamiltonian, Eq.~(\ref{hsec}). 
In the $(I,\hat{\theta})$ space, this Hamiltonian describes 
the motion in the vicinity 
of the fixed point at $I=n_0$. Expanding the Hamiltonian in
powers of $\tilde I=I-n_0$ around the fixed point yields
\be
{\cal H}_{\mathrm{sec}}\approx\hat{P}_t-\frac{3}{2n_0^2}
-\frac{3\tilde I^2}{2n_0^4}
-\frac{3F_sn_0^2}{2}\sqrt{1-\frac{L^2}{n_0^2}}\cos\psi
+F\Gamma(L,\psi)\cos[\hat{\theta}-\beta(L,\psi)].
\label{hsec2}
\ee
The explicit expressions for $\Gamma$ and $\beta$ are as follows
\be
\Gamma(L,\psi)=\sqrt{X_1^2\cos^2\psi+Y_{1}^2\sin^2\psi},
\ee
\be
\tan\beta(L,\psi)=\frac{Y_1}{X_1}\tan\psi,
\ee
where 
\begin{eqnarray} 
 X_{1}(n_0,L) = J'_1(e) n_0^{2} \\
 Y_1(n_0,L) = \frac{L}{n_0e} J_1(e) n_0^2
\end{eqnarray}
with $e=\sqrt{1-L^2/n_0^2}$ being
the eccentricity of the classical elliptical trajectory. $X_1$ and $Y_1$ are
nothing but the oscillatory atomic dipoles in resonance with the external
drive, along the major and minor axes of the classical Kepler ellipse,
respectively.

As there is no explicit time dependence in eq.~(\ref{hsec2}), 
the quantization of $\hat{P}_t$ is trivial \cite{szd99b,bsdz98}. 
Taking into account that Floquet eigenstates have to be periodic in time, this
yields $\hat{P}_t=k\omega$ (where $k$ is an integer number) which ensures the
periodicity of the quasienergy spectrum with a period $\omega$. 

The radial motion, i.e. in the $(\tilde I,\hat{\theta})$ space, 
effectively decouples 
from the slow angular 
motion in the $(L,\psi)$ space \cite{szd98}. In effect, one can quantize the
system in the spirit of the Born-Oppenheimer approximation, i.e. first quantize 
the radial motion keeping the secular motion frozen \cite{szd99b,bsdz98}
and then switch to the quantization of the slow $(L,\psi)$ motion. 
The radial motion reveals
a pendulum-like dynamics whose quantum eigenvalues are given by the solutions 
of the Mathieu equation \cite{abrammowitz72}. 
As we are looking for solutions with maximum localization inside the
resonance island, we will consider only
the ground state solution of the pendulum (excited states of the pendulum
describe the adjacent hydrogenic manifolds, see \cite{bsdz98}). We obtain:
\be
{\cal H}_{\mathrm{eff}}=-\frac{3}{2n_0^2}-\frac{3}{8n_0^4}a_0(q)-\frac{3F_sn_0^2}{2}
\sqrt{1-\frac{L^2}{n_0^2}}\cos\psi+k\omega,
\label{heff}
\ee
where
\be
q=\frac{4n_0^4F}{3}\Gamma(L,\psi)
\label{def_q}
\ee
is a dimensionless parameter. $a_0(q)$ is the Mathieu parameter corresponding to
the ground state of the pendulum \cite{abrammowitz72}.
The last stage is to quantize the secular motion which can be done directly 
using the WKB rule 
\cite{szd98,bsdz98}
\be
\frac{1}{2\pi}\oint Ld\psi=p+\frac{\mu}{4},   
\label{wkb}
\ee 
where $p$ is an integer number and $\mu$ stands for the Maslov index.

Without the static electric field, i.e. for $F_s=0$, it is more convenient 
to quantize the secular motion first 
(obtaining quantized values of $\Gamma$) and then switch to
quantization in the $(\tilde I,\hat\theta)$ space. This is allowed
because the entire dependence
of the Hamiltonian on $L$ and $\psi$ is included in $\Gamma(L,\psi)$
\cite{bsdz98}. In the presence of the static electric field, such a 
simplification is no longer
possible and one has to use the whole Hamiltonian (\ref{heff}) to perform
the semiclassical quantization in the $(L,\psi)$ space.

\begin{figure}
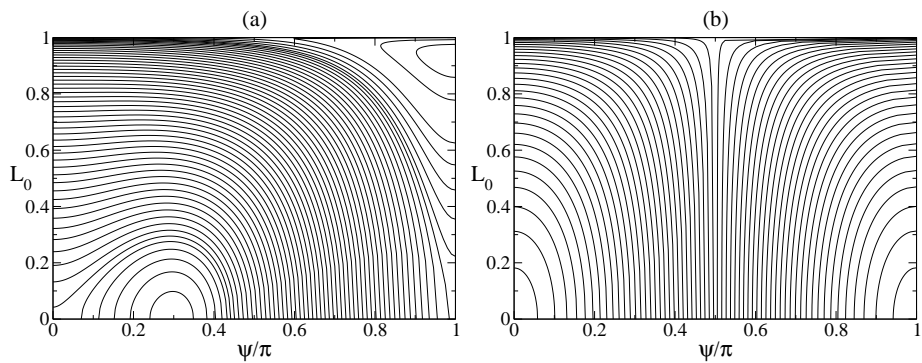

\begin{center}
\includegraphics*[width=6cm]{cont_a.eps}
\includegraphics*[width=6cm]{cont_b.eps}
\end{center}
\caption{
Structure of the $(L,\psi)$ phase space corresponding to 
the Hamiltonian ({\protect{\ref{heff}}}) for $n_0=60$, 
$F_{s,0}=n_0^4F_s=10^{-6}$ and for $F_{0}=n_0^4F=10^{-4}$ [panel (a)],
$F_{0}=n_0^4F=5\cdot 10^{-6}$ [panel (b)]. We use 
the scaled angular momentum 
$L_0=L/n_0$ in the plots. The ploted contours fulfill the semiclassical WKB
quantization rule given by eq.~({\protect \ref{wkb}}).
}
\label{phs1}
\end{figure}

Although, for comparison of the semiclassical quasienergies with the quantum
numerical values, we carry out calculations using the full Hamiltonian 
(\ref{heff}), it is instructive to perform further approximations and 
discuss weak and strong fields limit separately. 
For very small $F$ and moderate $n_0$, i.e. for $q\ll 1$, 
the Mathieu parameter can be approximated \cite{abrammowitz72} by
\be
a_0(q)\approx -\frac{q^2}{2}.
\ee
This is the regime corresponding to a very weak trapping pendulum potential, 
where the radial motion in the $(\tilde I,\hat{\theta})$ space is basically 
the free motion slightly 
perturbed (at second order in $F$) by the potential. However even for 
negligible external fields, the character of the secular motion is changed 
completely. That is, for $F=0$ and $F_s=0,$ 
both $L$ and $\psi$ are conserved quantities (i.e. the shape and the orientation 
of the electronic ellipse remain unchanged) while for $F\ne 0$ or $F_s\ne 0,$ 
the motion in the $(L,\psi)$ phase space may
reveal both librations (around a fixed point) and rotations as shown 
in Fig.~\ref{phs1}. Contours in Fig.~\ref{phs1} correspond actually to 
the semiclassically quantized states according to the WKB prescription, 
eq.~(\ref{wkb}), for $n_0=60$.
For fixed $F_s$ and with decreasing $F$, the fixed 
point on the $\psi=\pi$ axis moves from $L_0=L/n_0=1$ to 
$L_0=L/n_0=0$, see 
Fig.~\ref{phs1}. This
corresponds to an ellipse oriented along the field axis whose shape changes 
from a circle to a trajectory degenerated into a line. The eigenstate of the
system with the contour situated in the vicinity of 
this fixed point possesses probability
density localized around an ellipse whose eccentricity depends on $F/F_s$
ratio. 
However, there is no localization of an electron on such an elliptical  
trajectory because the
pendulum island in the $(\tilde I,\hat{\theta})$ space is too small to hold 
a semiclassical state --- the density probability is {\it equally} distributed 
along the ellipse with a weak (periodic) time dependence.

In the opposite limit, i.e. for large $F$ or in the deep semiclassical limit
$n_0\rightarrow\infty$, one may employ another asymptotic expression for
the Mathieu parameter \cite{abrammowitz72}
\be
a_0(q)\approx -2|q|+2\sqrt{|q|}.
\ee
This corresponds to the case where the pendulum, in the 
$(\tilde I,\hat{\theta})$ 
space, 
is localized near its stable equilibrium point. 
The equilibrium energy is $-2|q|$
while $2\sqrt{|q|}$ comes from 
the ground state energy of the pendulum calculated in 
the harmonic approximation. This is actually the approximation used in 
\cite{szd98}
where we have predicted the existence of  nondispersive wavepackets in 
this system.
The structure of the $(L,\psi)$ phase space has been presented in \cite{szd98} 
and 
is very similar to that shown for the weak fields limit --- e.g., 
there is also a fixed point located on 
the $\psi=\pi$ axis that
changes its position when $F/F_s$ varies. 
The state in the vicinity of this fixed point
corresponds to a well defined elliptical trajectory.
However, in the present case,  there is also localization of the electron on 
the trajectory because the island in the $(\tilde I,\hat{\theta})$ space is large 
enough 
to support quantum eigenstates. This allows us to build the nondispersive
wavepackets that are analyzed in the present article.


\end{document}